\documentclass[american, aps,pra, twocolumn, reprint ]{revtex4-1} 
\usepackage[unicode=true,pdfusetitle, bookmarks=true,bookmarksnumbered=false,bookmarksopen=false, breaklinks=false,pdfborder={0 0 0},backref=false,colorlinks=false] {hyperref}
\hypersetup{ colorlinks,linkcolor=myurlcolor,citecolor=myurlcolor,urlcolor=myurlcolor}
\usepackage{braket,colortbl,amsmath,cleveref,amsthm,amssymb,soul,xcolor}
\definecolor{myurlcolor}{rgb}{0,0,0.7}
\usepackage{graphics,graphicx}
\usepackage{color}
\usepackage{graphicx}
\usepackage[format = plain,labelfont = bf,up, textfont = normal , up, justification =raggedright, singlelinecheck =false]{caption}
\usepackage{subcaption}
\usepackage{tabularx} 
\DeclareMathOperator{\tr}{Tr}
\usepackage{graphicx}
\usepackage[utf8x]{inputenc}
\usepackage{color}
\usepackage{braket}
\usepackage{latexsym}
\usepackage{amssymb}
\usepackage{amsthm}
\usepackage{bm}
\usepackage{graphics,epstopdf}
\usepackage{color}
\usepackage{enumitem}
\usepackage{fmtcount}
\usepackage{booktabs}

\usepackage{epsfig}

\theoremstyle{plain}

\def\bea{\begin{eqnarray}}
\def\eea{\end{eqnarray}}
\def\ba{\begin{array}}
\def\ea{\end{array}}

\def\ket{\rangle}
\def\bra{\langle}
\def\beq{\begin{equation}}
\def\eeq{\end{equation}}

\begin{document}

\title{Wave-particle duality employing\\quantum coherence in superposition with 
non-orthogonal pointers}

\author{Sreetama Das}
\affiliation{Harish-Chandra Research Institute, HBNI, 
Chhatnag Road, Jhunsi, Allahabad 211 019, India}

\author{Chiranjib Mukhopadhyay}
\affiliation{Harish-Chandra Research Institute, HBNI, Chhatnag Road, Jhunsi, Allahabad 211 019, India}

\author{Sudipto Singha Roy}
\affiliation{Harish-Chandra Research Institute, HBNI, Chhatnag Road, Jhunsi, Allahabad 211 019, India}

\author{Samyadeb Bhattacharya}
\affiliation{Harish-Chandra Research Institute, HBNI, Chhatnag Road, Jhunsi, Allahabad 211 019, India}

\author{Aditi Sen(De)}
\affiliation{Harish-Chandra Research Institute, HBNI, Chhatnag Road, Jhunsi, Allahabad 211 019, India}
 
\author{Ujjwal Sen}
\affiliation{Harish-Chandra Research Institute, HBNI, Chhatnag Road, Jhunsi, Allahabad 211 019, India}

\begin{abstract}
\noindent  
We propose the notion of quantum coherence for superpositions over states
which are not necessarily mutually orthogonal. This
anticipatedly leads to a resource theory of non-orthogonal coherence. We
characterize free states and free operations in this theory, and connect
the latter with free operations in the resource theory of quantum
coherence for orthogonal bases. 
We show that the concept of non-orthogonal coherence naturally furnishes us with a wave-particle duality in quantum double-slit experiments where the 
channels beyond the slits are leaky between them.  
Furthermore, we demonstrate existence of a unique
maximally coherent qubit state corresponding to any given purity.
In addition, and in contradistinction with the case of orthogonal bases,
there appears a non-trivial minimally coherent qubit state for a given
purity. We also study the behavior of quantum coherence for some typical
configurations of non-orthogonal bases which have no analogs for
orthogonal bases. We further investigate the problem of determining the
energy cost of creating non-orthogonal coherence, and find that it scales
linearly with the non-orthogonal coherence created.

 
\end{abstract}

\maketitle

\section{Introduction}

The quantum superposition principle, rendering quantum mechanics distinct from
the laws of classical physics,  is one of the cornerstones in
the development of physics explaining phenomena around us and in laboratories. 
It is a crucial ingredient in a variety of
situations, ranging from Schr\"{o}dinger's thought experiment, known as
Schr\"{o}dinger's cat \cite{cat}, to the interference patterns in a quantum double-slit
experiment. It also leads to the phenomenon of 
non-distingushibility (i.e. non-discriminability) between pairs of quantum states, which is represented 
in the mathematical formalism by non-orthogonal states. The existence of non-orthogonal states puts constraints on certain tasks like copying \cite{nocloning}, deleting \cite{nodeleting}, and
broadcasting \cite{barnum} of unknown states, which are always possible in a classical world. In the last few decades, with the progress in quantum information, it has been realized
that the same linear superposition principle underlying such restrictions forms an essential element for understanding quantum communication and computational tasks \cite{wiesner,teleportation,shor,rsp2, rsp1}, including quantum cryptographic schemes \cite{BB84,e91,b92}.

Even though the linear superposition principle plays an important role in quantum
mechanics, in general, and quantum information, in particular, attempts to
systematically quantify the amount of superposition, termed  quantum
coherence, by using a fixed basis consisting of
orthonormal states, have been made rather recently \cite{aberg, baumgratz, winter}. Since its inception, quantum coherence has been linked with
other quantum resources like  non-locality \cite{debasis,nloc1,nloc2},
non-markovianity \cite{titas,nm1,nm2,nm3,nm4,nm5}, entanglement \cite{uttament,ent1,ent2,ent3,ent4,ent5,ent6,ent7, killoran}, and quantum discord
\cite{chinese,disc1,disc2}. Quantum coherence has also been shown to help in cooling \citep{huber} and work extraction \citep{anders,lostaglio} in quantum thermal machines, as well as augmenting quantum algorithms \citep{hillery, namit, hengfangrover}. Studies on the dynamics of coherence range from those in quantum spin models \citep{cakmak, spin1} and physical processes in biological systems \citep{bio1, bio3, bio2} to accelerated frames \citep{chiru2,hengfan} and curved spacetime \citep{curved}, pointing to the importance of quantum coherence over hugely different energy scales.  See Refs. \cite{cohrecent1,cohrecent2,cohrecent3,cohrecent4,cohrecent5} for most recent works on quantum coherence and other related concepts.

Quantification of quantum coherence of 
arbitrary quantum states  has until now been proposed in terms of fixed orthonormal bases. 
However, there is nothing sacred about orthonormal bases. One can just as well
write down a state in terms of a linearly independent basis, which spans the entire Hilbert space, but possibly includes mutually
non-orthogonal elements. Indeed, non-orthogonal bases naturally arise  in several areas of physics, ranging
from coherent states in quantum optics \citep{scullybook} to the Bennett 1992 \citep{b92} quantum key distribution
protocol.

It is worthwhile to mention here the connection of the theory of quantum coherence with pointers in measurement theory and devices. Measurement of a quantum state that is in a superposition of distinguishable physical situations (represented by orthogonal states in the corresponding Hilbert space) onto a basis of those orthogonal states can be interpreted in the measurement theory with these orthogonal states being identified with the pointers of the measuring device. Examples of such pointer states include slits in a two-slit experiment and paths in an interferometer. In the leaky two-slit experiment that we describe in the paper, the paths carry states that represent non-distinguishable physical situations (non-orthogonal states), and therefore may be thought of as corresponding to non-orthogonal pointers.

In this article, we have the following two aims. First, we 
attempt to construct a resource theory of quantum
coherence for arbitrary bases  by relaxing the
restriction of orthogonality. As in any resource theory, we define the free
states and free operations in this scenario. This is related to the similar concepts in the  resource theory of quantum coherence for orthonormal bases \citep{streltsov1, gour, streltsov2}. We then introduce a distance-based
measure of non-orthogonal coherence.  
After completion of the work in this manuscript, we were made aware of a
parallel work by Theurer \emph{et al.} \cite{plenio2}, also containing the concept
of non-orthogonal coherence and has a large coverage of facts
on the same, similar to the concepts we have introduced in this part of our work.
Second, we use this concept to obtain a wave-particle-type relation, in a quantum double-slit experiment 
with leaky post-slit channels, between the non-orthogonal coherence in the basis of the 
leaky channels and distinguishability of the detector states.

As further results, we find that in 
the case of qubits, and using a trace distance-based non-orthogonal coherence measure, there is a \emph{unique}  maximally coherent state corresponding to
any fixed non-orthogonal basis among states with a fixed amount of mixedness. This uniqueness is in 
sharp contrast to the usual family of maximally coherent mixed states
for orthonormal bases \citep{mcms}. We also demonstrate a trade-off between distance-based non-orthogonal quantum coherence measures and mixedness, the latter being quantified by linear or von Neumann entropy. Rather interestingly, and unlike the situation for orthonormal bases, we also witness the possible existence of states having non-zero \emph{minimal} non-orthogonal coherence for a given mixedness. Furthermore, we obtain lower and upper bounds on the sums of coherences with respect to different choices of non-orthogonal bases, including the ones that have no analogs in the orthonormal-bases scenario.  Moreover, we find that the energy cost \cite{samya} of generating non-orthogonal coherence is directly proportional to the non-orthogonal coherence created in the process. Therefore, we would like stress on the fact that though the  approach by Theurer \emph{et al.} \cite{plenio2} has a large coverage of facts
on non-orthogonal coherence, our manuscript contains independent
results on the energy cost of creating non-orthogonal coherence, wave-particle duality and
moreover, results highlighting the importance of the latter on
experimental interferometric setups.

 We wish to mention here that in Ref. \cite{Rastegin},  the author introduced coherence quantifiers, without the orthonormality restriction. However,  the reasoning there for   deviation from the orthogonality restriction was different from that in our case. 
 In Ref. Ref. \cite{Rastegin}, coherence was measured in the bases of an observable with degenerate eigenvalues. Hence, the notion of deviation from orthogonality was associated with the non-distinctness of the eigenvalues. The incoherent states arose as post-measurement states of the degenerate observable. For the non-degenerate case, the formalism coincided with the standard coherence theory. In addition to this, in Refs. \cite{Bischof1,Bischof2}, a proposal of resource theory of quantum coherence for general quantum measurements was discussed. Following the Naimark extension, the quantum states and operations were embedded into a higher-dimensional space. In this way, general measurements were extended to projective measurements for which the standard resource theory can be applied. Subsequently, free states and free operations were defined by analyzing the restriction to the embedded original space. Within our approach, we consider the resource theory of coherence while remaining in the same Hilbert space as the system. In this regard, it is interesting to mention the Gisin-Hughston–Jozsa–Wootters (GHJW) theorem \cite{Gisin,Hughston}, which states that arbitrary decompositions into pure states of any given mixed quantum state can be realized by measurements on an auxiliary system with which the original system is in an entangled pure state. In a theory of non-orthogonal coherence, one may however wish to consider measurements onto non-orthogonal bases that do not provide a decomposition of the state whose non-orthogonal coherence is sought. Indeed, it is the very existence of such pairs of bases and states that forms a motivation for this work.

The article is organized as follows. In Sec. \ref{anuchhed-dui}, we define free states,
free operations, and non-orthogonal quantum coherence measures. Sec. \ref{anuchhed-tin} 
is devoted to the derivations of non-orthogonal maximally and minimally coherent
states for arbitrary bases. 
The double-slit set-up and the corresponding duality is considered in Sec. \ref{anuchhed-char}.
A complementarity relation between non-orthogonal coherence and mixedness is provided in 
Sec. \ref{anuchhed-panch}. We also derive a non-trivial bound on the difference between mixedness and non-orthogonal coherence in the same section. In
Sec. \ref{anuchhed-chhoi}, we find lower and upper 
bounds on the total non-orthogonal coherence  with respect to more than one arbitrary bases arranged in specific configurations. We discuss the energy cost of generating non-orthogonal coherence in Sec. \ref{anuchhed-saath}. We present a conclusion in Sec. \ref{anuchhed-aat}.

\section{Free States and Free Operations}
\label{anuchhed-dui}

The formulation of a resource theory begins with the description of free states and free operations. We begin by introducing the free states in the resource theory of coherence for an arbitrary (possibly non-orthogonal) choice of basis. We refer to the free states as \emph{non-orthogonal incoherent states (NOIS)}, and define them as follows: 
\vspace{0.2 in}

\emph{\textbf{Definition:} If $\lbrace |a_{i} \rangle \rbrace_{i = 1,2,..,d}$ is an arbitrary basis (normalized but not necessarily mutually orthogonal) spanning the $d$-dimensional complex Hilbert space $\mathbb{C}_{d}$, then the family of states on $\mathbb{C}_{d}$ of the form \[ \chi = \sum_{i} p_{i} |a_{i} \rangle \langle a_{i}|  \] will be called non-orthogonal incoherent states (NOIS) for that basis, with $p_{i}$ being a probability distribution.} \\  Suppose now that in the definition of $\chi$, we allow the \(\{p_i\}\) to be a collection of complex numbers. The \(\{|a_i\rangle\}\) is still required to be a collection of linearly independent normalized vectors, and we also want \(\chi\) to be a density matrix. One can then show that among all collections of \emph{complex} numbers $\lbrace p_{i} \rbrace$, only probability distributions are allowed. The proof of this fact is as follows. Since $\lbrace |a_{i} \rangle \rbrace$ are linearly independent in $\mathbb{C}_{d}$, the projectors $\lbrace |a_{i}\ket\bra a_{i}| \rbrace$ are linearly independent in the space of bounded operators on $\mathbb{C}_{d}$. Then the hermiticity of $\chi$ implies $p_{i}^{*} = p_{i}$, so that $p_{i}$'s are real. The mixing weights $\lbrace p_{i} \rbrace$ must now be positive and have unit sum to keep $\chi$ within the space of density operators. Specializing to the case of a qubit ($d=2$), the NOIS are states on the straight line joining the two points on the surface of the Bloch sphere which represent the basis vectors. For a given basis, the non-NOIS form the ``non-free" or ``resourceful" states in this resource theory.

Free operations are those quantum operations by which no resource is created. For the resource theory of entanglement, these include the familiar local quantum operations and classical communication. For the resource theory of quantum coherence, these are the incoherent operations. We define our free operation as a generalization of the \emph{maximally incoherent operations (MIO)}, the largest class of incoherent operations known \citep{chitambar, streltsov}. We term them as \emph{non-orthogonal maximally incoherent operations (NOMIO)} and define them as follows.

 \vspace{0.1 in}

\emph{\textbf{Definition:} A quantum operation $\Lambda $ is said to be a non-orthogonal maximally incoherent operation (NOMIO) if it takes every NOIS to another NOIS.}

 \vspace{0.1 in}
 
It is natural to seek the connection between NOMIO and MIO. In the succeeding subsection, we show that a MIO sandwiched between two operations, each of which is what we will term as a \emph{basis changing (BC) operation}, can be a NOMIO.

\subsection{Characterization of NOMIO}

For two different orthonormal bases in a given state space, it is always possible to go from one to the other via a suitably chosen unitary operation. However, since unitary operations preserve the inner product between basis vectors, it is not possible to construct a  unitary operation $U$, on a single system, which maps an orthonormal basis $\lbrace |b_{i}\ket \rbrace$ to a normalized but non-orthogonal linearly independent basis $\lbrace |\tilde{b}_{i}\ket \rbrace$, that can be called  the \emph{target basis}. (By a non-orthogonal set of vectors, we will mean that at least one pair of vectors from the set is non-orthogonal.) If the magnitudes of inner products between all pairs of initial basis vectors are less than those between the corresponding pairs among the vectors in the target basis, we can overcome this difficulty, deterministically, by adding an auxiliary system (say, A) to the system denoted by S and performing a joint unitary operation on them followed by tracing out of the auxiliary system.   
  Such a representation would imply that the corresponding operation is physical. An explicit example of such operations is discussed later. 
We call such operations \emph{forward BC operations}. 

\emph{\textbf{Definition:} 
Suppose that there are two bases 
$B =\lbrace |b_{i}\ket\rbrace$ and $ C = \lbrace |c_{i} \ket\rbrace$, both containing normalized but not necessarily mutually orthogonal vectors, satisfying the condition $|\langle b_{i}|b_{j}\rangle| \leq |\langle c_{i}|c_{j}\rangle|$ \(\forall i,j\). The forward BC operation $L_{B \rightarrow C} $ is defined as} \beq L_{B \rightarrow C} (\rho_{S}) = \tr_{A} [U (\rho_{S} \otimes |0\ket_{A} \bra 0| ) U^{\dagger} ], \eeq 
\emph{ where the global unitary $U$ \cite{kraus} is chosen such that} \beq L_{B \rightarrow C} \left(|{b_{i}\ket}_{S}\bra b_{i}|\right) =   |c_{i}\ket_{S}\bra c_{i}| . \eeq

The unitary operator, $U$, is defined by
\begin{eqnarray}
U|b_i\rangle_s |0\rangle_A=|c_i\rangle_s |a_i\rangle_A, \forall i,\nonumber
\end{eqnarray}
where $\{|a_i\rangle\}$ form a set of normalized vectors for the Hilbert space corresponding to the auxiliary system. The fact that  such $|a_i\rangle$ does exist that allows $U$ to remain unitary follows from the relation below
\begin{eqnarray}
(_s\langle b_j| _A\langle 0|)  (|b_i\rangle_s |0\rangle_A)&=&(_s\langle c_j| _A\langle a_j|)  (|c_i\rangle_s |a_i\rangle_A),\nonumber\\ \text{i.e.,} \hspace{.1cm}
\langle b_j|b_i\rangle&=&\langle c_j|c_i\rangle \langle a_j|a_i \rangle,\nonumber
\end{eqnarray}
which, because of the condition $|\langle b_i|b_j\rangle|\leq |\langle c_i|c_j\rangle|, \forall i,j$, implies that $|\langle a_j|a_i \rangle |\leq 1$, which is generically valid for all pairs of vectors.

It is easy to see that the inverse operation of $L_{B \rightarrow C}$, that is, decreasing the inner product between bases cannot be carried out deterministically. Otherwise, one could take non-orthogonal basis vectors and perfectly distinguish their states by simply mapping this basis to an orthogonal basis. However, it is still possible to perform this operation, termed hereafter as the \emph{reverse BC operation} and denoted as $\tilde{L}_{C \rightarrow B}$, probabilistically, by using the unitary reduction map, similar e.g., to the technique used in probabilistic cloning machines \citep{duanpla,probclon1,probclon2}. The unitary-reduction procedure, in the present case, will be as follows. 
We will begin by implementing the unitary that transforms \(|c_i\rangle_S |0\rangle_A\) to 
\(\overline{a}_i |b_i\rangle_S |e_i\rangle_A + \overline{b}_i |\Phi_i\rangle_{SA}\), with the 
requirement that the span of the \(\{|e_i\rangle\}\) is orthogonal to the union of the supports of \(\{\mbox{tr}_S \left( |\Phi_i\rangle \langle \Phi_i | \right)_{SA}\}\). 
We then perform a measurement in the auxiliary system \(A\) that distinguishes between these orthogonal portions of the auxiliary Hilbert space, choose only the instances when the result of the measurement is onto the span of the \(\{|e_i\rangle\}\), and then trace out the auxiliary part.

Suppose now that we are interested in formulating a NOMIO for a non-orthogonal basis $C =  \lbrace |c_{i}\ket \rbrace $ in terms of some MIO $\tilde{\Lambda}_{B}$ for the choice of usual computational basis $B =  \lbrace |i\ket \rbrace $. There is no loss of generality, since any orthonormal basis can be reached from the basis, \(B\), via a unitary map. Clearly, the map from basis C to B is a reverse BC operation and the map from B to C is a forward BC operation.
 
\emph{\textbf{Proposition I:} The probabilistic quantum map  $ \Lambda  = L_{B \rightarrow C} \circ \tilde{\Lambda}_{B} \circ \tilde{L}_{C \rightarrow B} $ is a NOMIO for the basis $C$. }

\emph{Proof - } Suppose that an arbitrary NOIS for the basis $C$ is chosen as $\rho = \sum_{i} p_{i}|c_{i}\ket\bra c_{i}|$. Now $\Lambda (\rho) =  L_{B \rightarrow C}\circ \tilde{\Lambda}_{B} \circ \tilde{L}_{C \rightarrow B} (\sum_{i} p_{i}|c_{i}\ket\bra c_{i}|) $ = $L_{B \rightarrow C}\circ \tilde{\Lambda}_{B}  (\sum_{i} p_{i}|i\ket\bra i|)$ = $L_{B \rightarrow C} (\sum_{i} \tilde{p}_{i}|i\ket\bra i|)$ = $\sum_{i} \tilde{p}_{i}|c_{i}\ket\bra c_{i}| $ for some probability distribution $\lbrace \tilde{p}_{i} \rbrace$. This state is easily seen to be another NOIS, thus proving the  proposition. Here we have used the linearity of quantum operations as well as the MIO property of the map $\tilde{\Lambda}_{B}$. \qed

It is to be noted in this context that if the bases $B$ and $C$ are both orthonormal, then they are related through a unitary transformation and thus $L_{B\rightarrow C}$ simply equals $\tilde{L}_{C\rightarrow B}^{\dagger}$ and the reverse BC operation is also deterministic.

We also note here the case when the unitary-reduction map corresponding to the 
implementation of \(\tilde{L}_{C \rightarrow B}\) leads to a failure (measurement in the auxiliary part collapses to the union of the supports of \(\{\mbox{tr}_S \left( |\Phi_i\rangle \langle \Phi_i | \right)_{SA}\}\)), Proposition I does not indicate whether or not a ``cost" for the corresponding instances have occurred. If we do not wish to take into account the corresponding cost, the auxiliary required for the unambiguous distinguishing process and the state produced in case of a failure of the process can be declared as free states.

\subsection{Explicit example of a qubit forward BC operation}
Let us illustrate an example of a forward BC operation for a qubit system. The task is to transform the usual computational basis states $B = \lbrace |0\ket_{S} , |1\ket_{S} \rbrace $ to the non-orthogonal basis $ C =\lbrace |0\ket_{S} , |+ \ket_{S} = \frac{|0\ket_{S} + |1\ket_{S}}{\sqrt{2}} \rbrace $.

We first add an auxiliary  qubit $|0\ket_{A}$ and perform the following operations in sequence.
\begin{enumerate}
\item Execute a controlled Hadamard gate which takes $|0\ket_{S} |0\ket_{A} \longrightarrow  |0\ket_{S} |0\ket_{A} $ and $|1\ket_{S} |0\ket_{A} \longrightarrow  |1\ket_{S} \otimes U^{H} |0\ket_{A} = |1\ket_{S} |+\ket_{A} $, where $U^{H}$ is the Hadamard gate \citep{nielsenbook}.

\item Now apply the swap gate which takes  $|0\ket_{S} |0\ket_{A} \longrightarrow  |0\ket_{S} |0\ket_{A} \ket $ and $|1\ket_{S} |+\ket_{A} \longrightarrow  |+\ket_{S} |1\ket_{A} $.

\item Finally, trace out the auxiliary part. 

\end{enumerate}

In this way, using global unitary operations on the system-auxiliary pair, the orthogonal basis B can be changed to the non-orthogonal basis C in the system.  This implies that the required operation to change the basis is physically implementable.  We  mention here that the scheme described above   is not unique, and in principle, there could be other possible  methods of implementations of  the forward BC operation

\subsection{Towards non-orthogonal incoherent operations}
Maximally incoherent operations form the largest set of free operations in the resource theory of coherence. 
A more restricted class of free operations which play an important role in the resource theory of quantum coherence  are the \emph{incoherent operations (IO)}  \citep{baumgratz}.


In analogy with IO, we define a \emph{non-orthogonal incoherent operation (NIO)} as any channel $\Lambda_{NIO}$ with Kraus operators $\lbrace K_{i} \rbrace$ such that $K_{i}\chi K_{i}^{\dagger}/ \tr(K_{i}\chi K_{i}^{\dagger}) \in \lbrace \chi \rbrace$ $\forall \chi \in \lbrace \chi \rbrace $, with $\lbrace \chi \rbrace $ being the set of NOIS.

\subsection{Quantification of non-orthogonal coherence}

Formulation of free states and operations sets the stage for any resource theory. An immediate next step is to quantify the amount of resource a state possesses. 
Towards this aim, we propose the following geometric measure of non-orthogonal coherence. 

\emph{\textbf{Definition:} For any quantum state $\rho$, a bona fide measure of non-orthogonal coherence $C^{NO}$ is given by   
\begin{equation}
\label{cno}
C^{NO} (\rho) = \min_{\chi} D\left(\rho, \chi\right),
\end{equation} 
where D is a suitable contractive distance measure and the minimization is over all NOIS for that particular basis.} 

Confining ourselves to the qubit scenario and the usual Bloch sphere representation, we note that if we take this distance measure as the trace distance, it corresponds to the Euclidean distance for the Bloch sphere. Since trace distance was shown to be a valid coherence measure for qubits in the case of orthonormal bases \citep{swapan}, we choose our bona fide non-orthogonal coherence measure in terms of the trace distance as 
\begin{equation}
\label{cnotr}
C^{NO}_{trace} (\rho) = \min_{\chi} \tr|\rho - \chi|.
\end{equation}  
In the Bloch sphere representation, it corresponds to the shortest distance of a point from the line joining the non-orthogonal basis vectors.

Let us now enumerate  the necessary requirements that we intuitively expect such a coherence measure $C^{NO}$ to satisfy.

\begin{enumerate}[label={(\bfseries P\arabic*)}]
\item $C^{NO} \left(\chi\right) = 0$ for any NOIS  $\chi$, and positive otherwise.

\item  $C^{NO} \left(\Lambda_{NIO}(\rho)\right) \leq C^{NO} \left(\rho\right)$ for any state $\rho$, and an arbitrary $\Lambda_{NIO}$.

\end{enumerate}

The properties of any contractive distance measure guarantees the fulfillment of (P1) and (P2). Other non-orthogonal coherence measures can also be obtained by using the Renyi or Tsallis family of relative entropies \citep{rastegin, renyicoh}, although we remember that relative entropy-based distance measures do not satisfy the symmetry property and the triangle inequality.

 It is not clear if $C^{NO}$ would satisfy a postulate on monotonicity with respect to selective measurements and on convexity.

\section{Maximally coherent States for Non-Orthogonal Basis}
\label{anuchhed-tin}

In this section, we search for states which possess the maximal amount of non-orthogonal coherence. In case of the resource theories of coherence and entanglement, there exists a family of infinitely many maximally resourceful states interconnected via unitaries. However, as we show below, this is no more true for coherence measures based on a non-orthogonal basis. 

\begin{figure}
\includegraphics[scale=0.3]{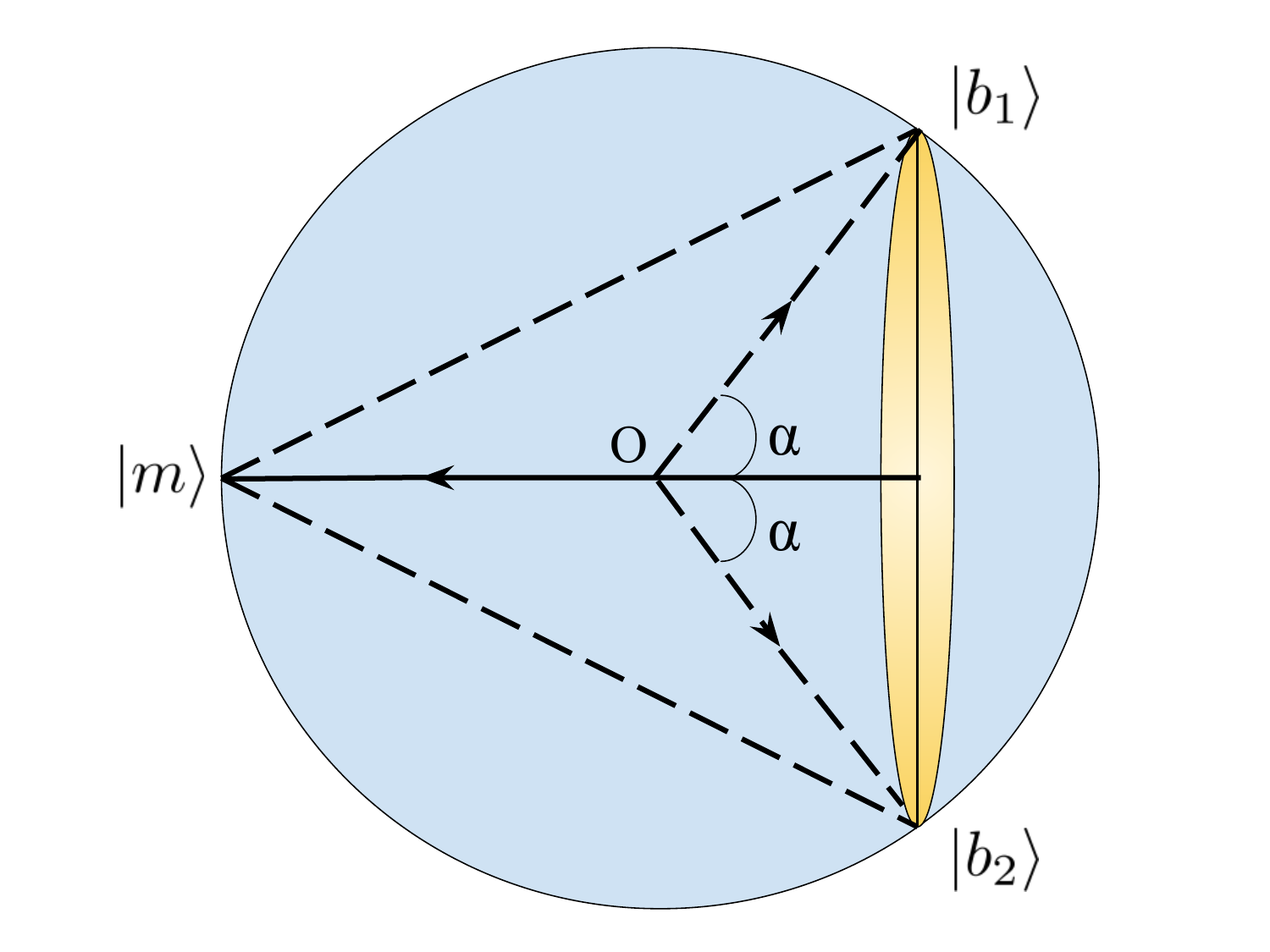}
\caption{(Color online) Pictorial Bloch sphere representation of maximally coherent qubit state $|m\ket$ for the non-orthogonal basis vectors $\lbrace |b_{1}\ket, |b_{2}\ket \rbrace$} 
\label{fig1}
\end{figure}

We will call a state as maximally coherent for a given dimension if \(C^{NO}_{trace}\) 
attains the maximal value for that state for that dimension.

\emph{\textbf{Proposition II:} If $\lbrace |b_{1}\ket, |b_{2}\ket \rbrace$ constitute a non-orthogonal basis, then in this basis, the measure based on the trace norm, $C_{trace}^{NO}$, is maximized for a unique pure state given by $|m\ket\bra m| = \frac{1+ \cos \alpha}{\cos \alpha} \frac{\mathbb{I}}{2} - \frac{1}{2 \cos \alpha} \left( |b_{1}\ket \bra b_{1}| + |b_{2} \ket \bra b_{2}| \right)$, where $2\alpha$ is the angle subtended at the origin by the points on the Bloch sphere that represent $|b_{1}\ket$ and $|b_{2}\ket$, and where $\mathbb{I}$ is the identity operator on the qubit Hilbert space.}                                                                                                                                                                                                                                                                                                                                                                                                                                                                                                                                                                                                                                                                                                                                                                       

\emph{Proof -} Consider the pure state $|m\ket\bra m|$ on the line joining the center of the Bloch sphere and the midpoint of the line joining the basis vectors (see Fig. \ref{fig1}). It can be easily shown that of all the point in the Bloch sphere, $|m\ket$ is the farthest  from the line joining $|b_{1}\ket$ and $|b_{2} \ket$ in terms of the usual Euclidean distance. Since the trace distance formula reduces to the Euclidean distance in case of Bloch spheres representing qubit states, $|m\ket$ is indeed the maximally coherent state with respect to the non-orthogonal basis. \qed

A pictorial representation of the maximally coherent state is given in Fig. \ref{fig1}. In Table I, we provide a comparison between a few well-known resource theories in the 
quantum world in terms of their respective free states and maximally resourceful states.

\begin{table}[htb]
\centering
\begin{tabular}{ | l | c  | c|}
    \toprule 
    \hline  
    \textbf{Quantum resource theory}  & \textbf{Free states} & \textbf{Maximally} \\
     & & \textbf{resourceful states} \\
    \hline 
    \midrule  
    Entanglement  & $\infty$   &  $\infty$   \\  
  \hline
    Coherence & $\infty$   &  $\infty$   \\  
  \hline
  Thermodynamics &  &     \\
  (for states diagonal & 1 (thermal state) & 1 (fully excited state)\\
   in energy eigenbasis) & & \\
  \hline
  Non-orthogonal coherence  & $\infty$   &  1 (in the qubit case) \\
  \hline
    \bottomrule
\end{tabular}
\caption{Comparing the well-known quantum resource theories of entanglement 
\citep{horodeckireview}, coherence \citep{adessorev}, and thermodynamics \citep{nanoscale, brandao} with the resource theory of non-orthogonal coherence in terms of the number of free and maximally resourceful states allowed.}
\end{table}

\section{Leaky quantum double-slit set-up and a complementarity}
\label{anuchhed-char}
\begin{figure}
\hspace{-1.5cm}
\includegraphics[width=.45\textwidth, angle =0] {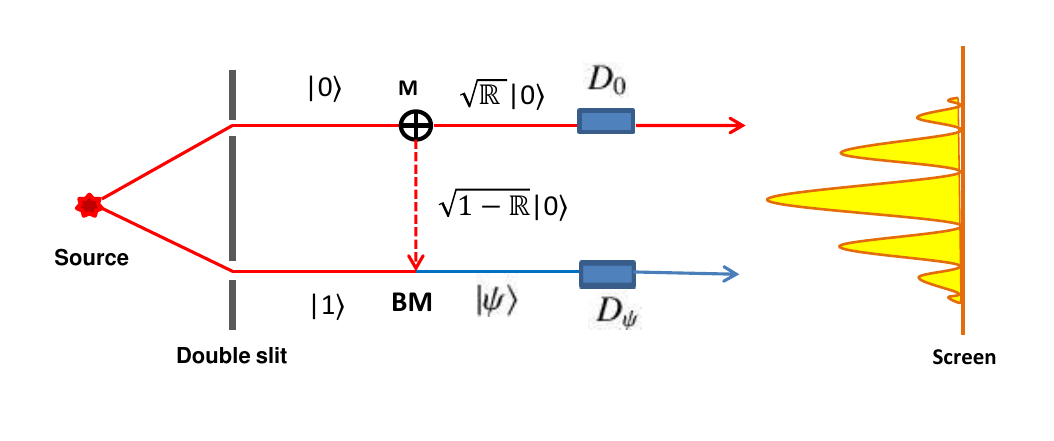}
\caption{(Color online) Schematic diagram of the quantum double-slit experiment with a ``leak''.
A source emits a quantum system \(\mathbb{Q}\) that goes through two slits. The leak, between the beams from the two slits, is modelled by a device M that lets the incoming beam pass with a certain probability, and coherently diverts to the other beam with the remaining probability. A phase shifter (not shown in the diagram) is inserted in the diverted channel to compensate for the accumulated phase due to the extra path travelled by that beam. To extract a partial knowledge about the particle nature
of the system \(\mathbb{Q}\), a unitary-reduction process is effected, and is denoted by the devices
\(D_0\) and \(D_\psi\). See text for further details.} 
\label{fig-double-slit}
\end{figure}

In this section, we will use the proposed measure
of non-orthogonal coherence to uncover a wave-particle-like complementarity, in a
leaky quantum two-slit set-up armed with partial detectors, between coherence of the state
that reaches the screen and the path-distinguishability obtained via the detectors. The wave-particle duality is an important
trait of quantum mechanics and has been seen from several perspectives.
The majority of these applications are however for ideal non-leaky
set-ups. Even though there have been sustained improvements in checking
the relations in experiments, the set-ups are almost always non-ideal. We
show that the concept of non-orthogonal coherence naturally furnishes us
with a wave-particle duality in quantum double-slit experiments where the
channels beyond the slits are leaky between them. This provides an
operational meaning to non-orthogonal coherence in a realistic set-up.
More precisely, we consider the set-up 
of the usual double-slit experiment \cite{double_slit_book},
 with the states immediately after the two slits, ``upper''
and ``lower'', being
denoted by \(|0\rangle\) and \(|1\rangle\), respectively. However, the channels carrying the
two states are, in our case, ``leaky'', and the state of one is affected by the other. This 
is schematically described in Fig. \ref{fig-double-slit}. 
We model the ``leak''
by inserting a
device, M,
close to the upper slit. The device M
lets the state of that channel to go through with probability \(\mathbb{R}\), while leaking it into the other channel with the remaining probability. We also insert a phase shifter (not shown in the
schematic diagram) in the diverted channel to annihilate the effect of the accumulated phase
due to the extra path traversed in the diverted channel.
The entire process is, for simplicity, assumed to be coherent, and
in particular, the environment does not additionally disturb the process.
However, by continuity, the results obtained
will remain near-optimal even for weak environmental noise.
Note that the mathematics of
the process is equivalent to a Mach-Zehnder
interferometric set-up with similar leaky channels.
Right after the device M, and still close to the slits, we put probes (``detectors'') \(D_0\)  and \(D_\psi\)
represented by states \(|d_0\rangle\) and \(|d_\psi\rangle\) respectively, 
to obtain (partial) information about
which slit the quantum state came out of.
Just after the device M and the phase shifter, but before the detectors are inserted,
the quantum state of the system would be
\begin{eqnarray}
|\Psi\rangle_{tot}=(\alpha \sqrt{\mathbb{R}} |0\rangle + \sqrt{M}|\psi\rangle)/\sqrt{N},
\end{eqnarray}
 where
 \begin{eqnarray}
 |\psi\rangle = (\beta |1\rangle + \alpha \sqrt{1 - \mathbb{R}}|0\rangle)/\sqrt{M},
 \end{eqnarray}
\(M = |\beta|^2 + |\alpha|^2 (1-\mathbb{R})\),
\(N=1 + 2|\alpha|^2 \sqrt{\mathbb{R}(1-\mathbb{R})}\), where
the input state is \(\alpha |0\rangle + \beta |1\rangle\), 
\(\alpha\) and \(\beta\) being arbitrary complex numbers with \(|\alpha|^2 + |\beta|^2 = 1\).
The state \(|\psi\rangle\) is the quantum state of the channel from the lower slit, just after the diverted
channel joins it. The joining point is denoted in the figure as a ``beam merger'' (BM).

After insertion of the detectors (i.e., after the system and the detectors have interacted), which is followed by a unitary-reduction process (see below),
the state of the joint quantum system, consisting of the system that
passes through the slits (call it \(\mathbb{Q}\)) and that of the detectors (call it \(\mathbb{D}\)),
can be represented as
\begin{eqnarray}
|\Psi\rangle_{\mathbb{QD}} = (\alpha \sqrt{\mathbb{R}} |0\rangle |d_0\rangle +
\sqrt{M}|\psi\rangle |d_\psi\rangle)/\sqrt{R},
\end{eqnarray}
 where \(R\) is a normalization constant given by
\(R= \frac{|\alpha|^2 \mathbb{R}}{N} + \frac{M}{N} + \frac{2|\alpha|^2 \sqrt{M \mathbb{R} (1-\mathbb{R})}}{N} \Re[\langle d_\psi | d_0\rangle]\). (The detector states are not necessarily orthogonal.)
This is to be compared
with the ``quanton-detector'' state in the landmark paper of Englert \cite{englert1} (see also
\cite{GREENBERGER1988391,Jaeger-shimony, Zurek, Englert1992, Mandel:91, Manab}),
which considered the same experiment, but without the device M (and the corresponding phase shifter).
Note that the insertion of the detectors in the current set-up requires a
unitary-reduction process (see e.g. \cite{DuanGuoPRL}, \cite{duanpla}), in contrast to the unitary process in \cite{englert1}.
The unitary-reduction process by which the probes in system \(\mathbb{D}\) interact with the system \(\mathbb{Q}\) works as follows: It begins with a unitary process that executes
the transformation, given by 
\begin{eqnarray}
|0\rangle_{\mathbb{Q}}|d\rangle_{\mathbb{D}} &&\rightarrow
\sqrt{q} |0\rangle_{\mathbb{Q}} |d_0 \rangle + \sqrt{1-q} |\Psi{'}\rangle_{\mathbb{QD}},\nonumber\\
|\psi\rangle_{\mathbb{Q}}|d\rangle_{\mathbb{D}}&& \rightarrow
\sqrt{q} |\psi\rangle_{\mathbb{Q}} |d_\psi\rangle + \sqrt{1-q} |\Psi{''}\rangle_{\mathbb{QD}}. 
\end{eqnarray}
The states
\(|\Psi{'}\rangle\) and \(|\Psi{''}\rangle\) are so chosen that inner product on the left- and right-hand
sides coincide. Moreover, we also ensure that the span of \(\{|0\rangle|d_0\rangle, |\psi\rangle |d_\psi\rangle\}\) is orthogonal to that of \(\{|\Psi{'}\rangle, |\Psi{''}\rangle\}\). Here, \(q\) is a probability that is maximized under the constraints of unitarity and orthogonality of the spans.
The unitary
interaction is followed by a measurement distinguishing between these two spans, and we choose only
those instances when the joint system is projected onto the span of \(\{|0\rangle|d_0\rangle, |\psi\rangle |d_\psi\rangle\}\).
The local density matrix of the system \(\mathbb{Q}\) is given by
\begin{eqnarray}
\varrho_{\mathbb{Q}} = \mbox{tr}_{\mathbb{D}} (|\Psi\rangle\langle\Psi|)_{\mathbb{QD}},
\end{eqnarray}
 and
we consider its coherence \(C^{NO}_{trace} (\varrho_{\mathbb{Q}})\) in the non-orthogonal
basis \(\{|0\rangle, |\psi\rangle\}\).
Let \(|m\rangle\) denote the state which possesses the maximal \(C^{NO}_{trace}\) for the non-orthogonal basis \(\{|0\rangle, |\psi\rangle\}\).
The normalized non-orthogonal coherence, 
\begin{eqnarray}
\widetilde{\mathcal{C}}=C^{NO}_{trace} (\varrho_{\mathbb{Q}})/C^{NO}_{trace}(|m\rangle),
\end{eqnarray}
is identified as the residual
wave nature present in the system \(\mathbb{Q}\), even after its interaction with the
detectors.
From the state of the other half of the \(\mathbb{QD}\) system, we can quantify the
particle nature extracted from the system \(\mathbb{Q}\) by using the probes.
Tracing out the system \(\mathbb{Q}\), we obtain
\begin{eqnarray}
\varrho_{\mathbb{D}} = \mbox{tr}_{\mathbb{Q}} (|\Psi\rangle\langle\Psi|)_{\mathbb{QD}},
\end{eqnarray}
 which we
phase damp in the (non-orthogonal) basis \(\{|d_0\rangle, |d_\psi\rangle\}\) to obtain
\begin{eqnarray}
\varrho{'}_{\mathbb{D}} = |\alpha|^2 \mathbb{R} |d_0\rangle \langle d_0|
+ (1-|\alpha|^2 \mathbb{R}) |d_\psi\rangle \langle d_\psi|. 
\end{eqnarray}

To phase damp in the non-orthogonal basis, \(\{|d_0\rangle, |d_\psi\rangle\}\), we begin by 
performing a positive operator valued measurement with the elements 
\begin{eqnarray}
A_0 &=& c(\mathbb{I}-|d_\psi\rangle \langle d_\psi |,\nonumber \\
A_\psi &=& c(\mathbb{I}-|d_0\rangle \langle d_0 |),\nonumber\\
A_? &=& \mathbb{I} - A_0 - A_\psi.
\end{eqnarray}
 The real number \(c\) is chosen to be the maximal positive number so that \(A_?\) is positive semi-definite. This is exactly the same measurement required for unambiguous 
discrimination of two non-orthogonal states \cite{IVANOVIC1987257,CHEFLES1998339,0305-4470-31-50-007,Barnett:09, Dieks, Jaeger, Peres}, required, e.g., for the 
Bennett 1992 quantum cryptography protocol \cite{b92}. For an input \(\rho\) to this measurement, 
we post-select the results that does \emph{not} click in \(A_?\), and also forget the distinction between 
the instances when the other two have clicked, so that the (unnormalized) output, within the von Neumann-L{\"u}ders 
 postulate \cite{kraus}, is 
 \begin{eqnarray}
  A_0^{1/2} \rho A_0^{1/2} + A_\psi^{1/2} \rho A_\psi^{1/2}.
 \end{eqnarray}
 This state is then rotated by a unitary 
 that realizes \(|d_\psi^\perp\rangle \to |d_0\rangle\) and \(|d_0^\perp\rangle \to |d_\psi\rangle\), 
 where \(\langle d_0|d_0^\perp \rangle = 0 = \langle d_\psi|d_\psi^\perp \rangle \).

The maximal probability for unambiguous discrimination between
the non-orthogonal states \(|d_0\rangle\) and \(|d_\psi\rangle\), generated with probabilities
\(|\alpha|^2 \mathbb{R}\) and \(1 - |\alpha|^2 \mathbb{R}\) respectively, is given by \cite{IVANOVIC1987257,CHEFLES1998339,0305-4470-31-50-007,Barnett:09, Dieks, Jaeger, Peres}
\begin{eqnarray}
\widetilde{\mathcal{D}} = 1- 2\sqrt{|\alpha|^2 \mathbb{R}(1-|\alpha|^2 \mathbb{R})} \left|\langle d_0|d_\psi\rangle\right|. 
\end{eqnarray}
The quantity \(\widetilde{\mathcal{D}}\) describes the  particle nature
that has been extracted from the system \(\mathbb{Q}\) by using the detectors.
\begin{figure}
\hspace{0cm}
\includegraphics[width=.40\textwidth, angle =0] {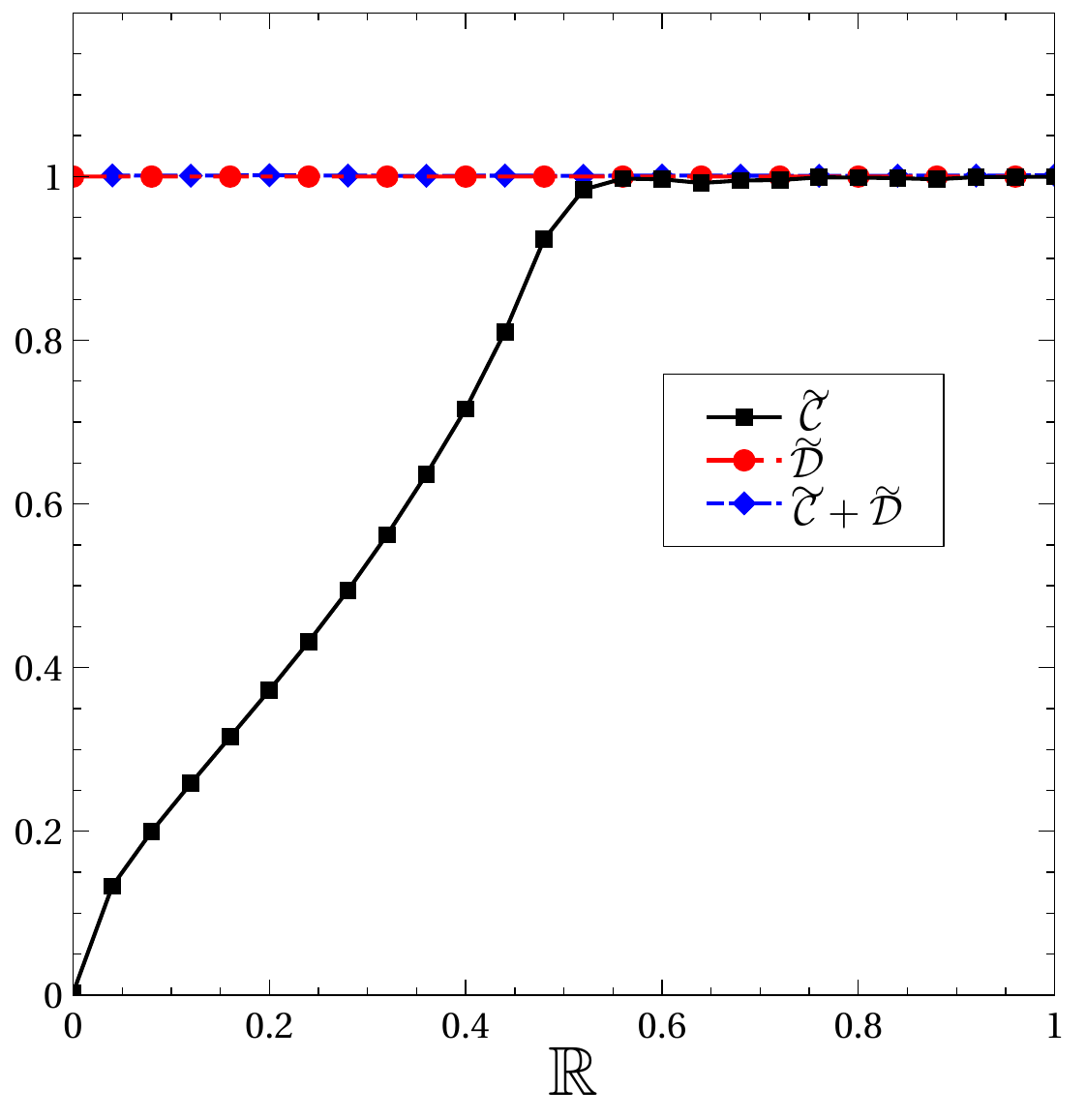}
\caption{(Color online) A wave-particle duality in a quantum double-slit set-up with a leak. For the curve with blue diamonds, the vertical axis represents the maximal value of the sum
\(\widetilde{\mathcal{C}} + \widetilde{\mathcal{D}}\) for a given \(\mathbb{R}\), with the latter being on the horizontal axis. For the curve with black squares (red circles), the vertical axes represents the maximal value of the quantity \(\widetilde{\mathcal{C}}\) (\(\widetilde{\mathcal{D}}\)) on the same horizontal axis. For a given \(\mathbb{R}\), all quantities on the  vertical axis are maximized over the entire parameter space. Note that the curves for  \(\widetilde{\mathcal{C}}\) and \(\widetilde{\mathcal{D}}\) do not add up to that for  \(\widetilde{\mathcal{C}} + \widetilde{\mathcal{D}}\). Note also that the curves for \(\widetilde{\mathcal{C}} + \widetilde{\mathcal{D}}\) and for \(\widetilde{\mathcal{D}}\) have merged with each other. All quantities are dimensionless.
}
\label{chhabi-India-marching-for-science}
\end{figure}

The normalized non-orthogonal coherence, \(\widetilde{\mathcal{C}}\), as well as the maximal probability for unambiguous distinguishability, \(\widetilde{\mathcal{D}}\), are individually
bounded above by unity. However, by Haar uniformly generating \(10^6\) points in the particle-detector parameter space, we find that the sum of  $\widetilde{\mathcal{C}}$  and $\widetilde{\mathcal{D}}$, maximized over all these $10^{6}$ points remains at unity, independent of the leakage amount:
\begin{equation}
\widetilde{\mathcal{C}} + \widetilde{\mathcal{D}} \leq 1.
\end{equation}
Note that the above bound is 
\(\mathbb{R}\)-independent. See Fig. \ref{chhabi-India-marching-for-science}. Note also that
 \(\mathbb{R}=1\) implies that
the device D is absent from the set-up, in which case, our results reproduce a bound obtained in Ref.
\cite{Manab}.

\begin{figure}
\includegraphics[width=.40\textwidth, angle =0] {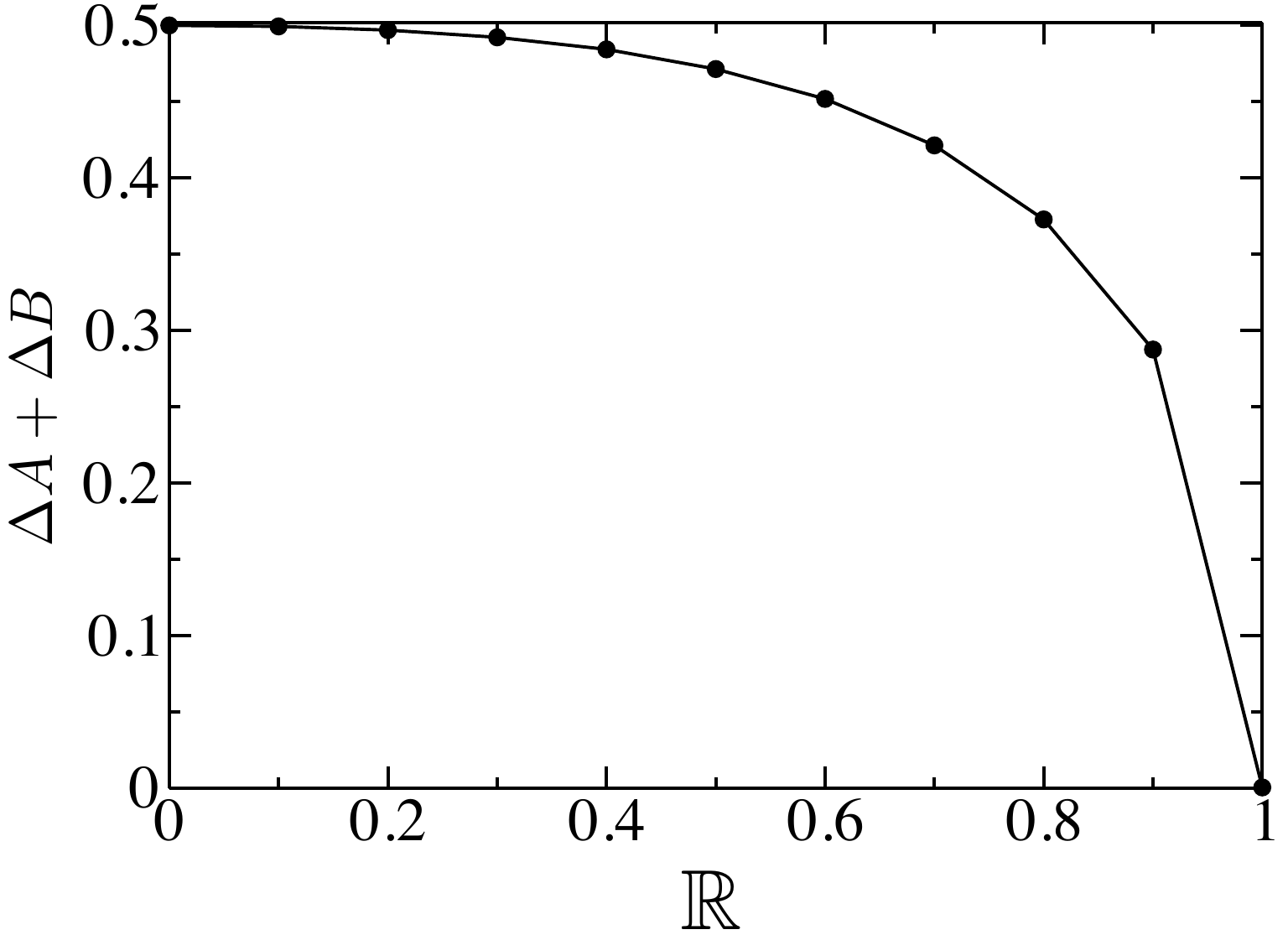}
\caption{Minimal uncertainty sum for observables in the leaky double-slit set-up.
We plot here the minimum value of the quantity \(\Delta A + \Delta B\),
for a fixed input state to the double-slit set-up, but minimized over all
states \(|\phi\rangle\), against \(\mathbb{R}\). The minimization is
performed by Haar uniform generation of \(10^7\) system states. All
quantities are dimensionless.}
\label{fig:uncertainty}
\end{figure}

 It is interesting to consider the quantum uncertainty between noncommuting
observables \cite{sakuraibook} that naturally appear in the leaky double-slit experiment
considered here. The detectors in the set-up in Fig. 2 tries to detect
whether the superposition \(|\Psi\rangle_{tot}\) is in the state
\(|0\rangle\) or in  \(|\psi\rangle\). This naturally leads us to the
operators \(A=|0\rangle \langle 0|\) and \(B = |\psi\rangle \langle
\psi|\). \(A\) and \(B\) are functions of the parameters in
\(|\psi\rangle\), and become commuting when \(|\psi\rangle\) becomes equal
or orthogonal to \(|0\rangle\). For definiteness, we consider the case
when $\alpha = \beta = \frac{1}{\sqrt{2}}$. For an arbitrary qubit state
\(|\phi\rangle\), the standard deviation, \(\Delta A\), of \(A\) is given
by the positive square root of $\langle \phi| A^2 |\phi\rangle - \langle
\phi | A |\phi\rangle^2$. \(\Delta B\) is similarly defined. We then
minimize the quantity \(\Delta A + \Delta B\) \cite{Pati,Sun,Fei,Xue,Maziero} over the states
\(|\phi\rangle\), for a fixed \(\mathbb{R}\). The behavior of the
uncertainty sum is then plotted with respect to \(\mathbb{R}\) in Fig. \ref{fig:uncertainty}. We find that although the minimal uncertainty sum has significant
variation for different values of the parameter \(\mathbb{R}\), the
maximal wave-particle duality sum, \(\tilde{\mathcal{C}} +
\tilde{\mathcal{D}}\) remains frozen in that range.  Note here that in general, a lower bound on $\Delta A+ \Delta B$
$(\Delta A, \Delta B \ge 0)$ will not provide a nontrivial (i.e., nonzero) lower bound on $\Delta A \Delta B$. Even in the particular case at hand, an optimization over $|\phi\rangle$ will
take the minimum of $\Delta A \Delta B$ to zero, ruling out a nontrivial
state-independent lower bound on $\Delta A \Delta B$.

\section{Non-orthogonal coherence vs. mixedness}
\label{anuchhed-panch}

The concept of maximally resourceful mixed states, i.e., the states possessing maximal amount of a resource given a certain mixedness, have been introduced in resource theories of entanglement  and coherence, known respectively as \emph{maximally entangled mixed states (MEMS)} \citep{mems} and \emph{maximally coherent mixed states (MCMS)} \citep{mcms}. In each of these cases, the MEMS or the MCMS are an infinite set for a given mixedness. However, from Fig.\ref{fig1}, it is easy to see that this is again not the case for quantum coherence in a non-orthogonal basis in the qubit scenario. The radial distance of a point inside the Bloch sphere from the origin is a measure of purity of the state represented by that point.  The complement to this measure of purity, known as the linear entropy measure of mixedness, is clearly equal on all points lying on the surface of a sphere concentric to the Bloch sphere and of radius $0 < r \leq 1$. Thus we can parametrize all points lying on the surface of this sphere as having the same amount of mixedness. Now it can be shown, just like the pure state case, that the state on this sphere which is on the extended straight line joining the origin and the midpoint of the line segment between the basis vectors, is the one with maximal non-orthogonal coherence. This state may be written as $\rho_{max} = \frac{r+ \cos \alpha}{\cos \alpha} \frac{\mathbb{I}}{2} - \frac{r}{2 \cos \alpha} \left( |b_{1}\ket \bra b_{1}| + |b_{2} \ket \bra b_{2}| \right)$, where r is the radius of this sphere, and $\alpha$ is defined as in the enunciation of Proposition II.  Now, since we can take $1-r$ as a measure of mixedness $M$, we note that the maximally resourceful state $\rho_{max}$ has non-orthogonal coherence $C_{trace}^{NO} (\rho_{max}) = r+ \cos \alpha = 1+ \cos \alpha - M(\rho_{max})$. Therefore, for an arbitrary state $\rho$ with mixedness $M$, we have \[ C_{trace}^{NO} (\rho) \leq r+ \cos \alpha = 1+ \cos \alpha - M(\rho), \] which leads to the complementarity relation 
\begin{equation}
\label{nmcms}
C_{trace}^{NO} (\rho)+ M(\rho) \leq 1+ \cos \alpha, 
\end{equation}for all quantum states $\rho$. The corresponding inequality in the case of an orthogonal basis was given in \citep{mcms}.

However, it can be shown that mixed states with  a fixed amount of purity, parametrized by the spherical surface with Bloch radius $r$, if $r$ is less than a \emph{purity threshold} \beq{\tilde{r} = \cos \alpha, }\eeq cannot be incoherent, as the spherical surface then does not intersect the line of incoherent states. In other words, we have the following ``no-go''-like result.

\emph{\textbf{Proposition III:} For any pair of non-orthogonal basis vectors of a qubit that has an inner product \(\cos\alpha\), there are no mixed states, \(\rho\), that are incoherent if \(\mbox{Tr} (\rho^2)\) $<$ \(\frac{1}{2}(1+\cos^2\alpha)\).}

The antipodal point to the non-orthogonal MCMS (NOMCMS) possesses the minimum trace distance-based coherence $C^{NO}_{trace} = \cos \alpha - r $ among all states having the same purity \(r< \tilde{r} \). This state may thus be called the unique  \emph{non-orthogonal minimally coherent mixed states (NOMinCMS)} and consequently for an arbitrary state $\rho$, we have

\begin{equation}
\label{mincms1}
C_{trace}^{NO}(\rho) \geq \cos \alpha - r = \cos \alpha - 1 + M(\rho)),
\end{equation}  

\noindent which gives an additional relation between mixedness and non-orthogonal coherence measure, which is trivial for quantum coherence with orthogonal bases:

\begin{equation}
\label{mincms2}
 M(\rho) - C_{trace}^{NO}(\rho) \leq  1 - \cos \alpha.
\end{equation}
This relation shows us how the possibility of defining quantum coherence with respect to a non-orthogonal basis leads us to qualitatively different scenarios and results compared to the case of orthonormal bases. 

\begin{figure}
\includegraphics[scale=0.5]{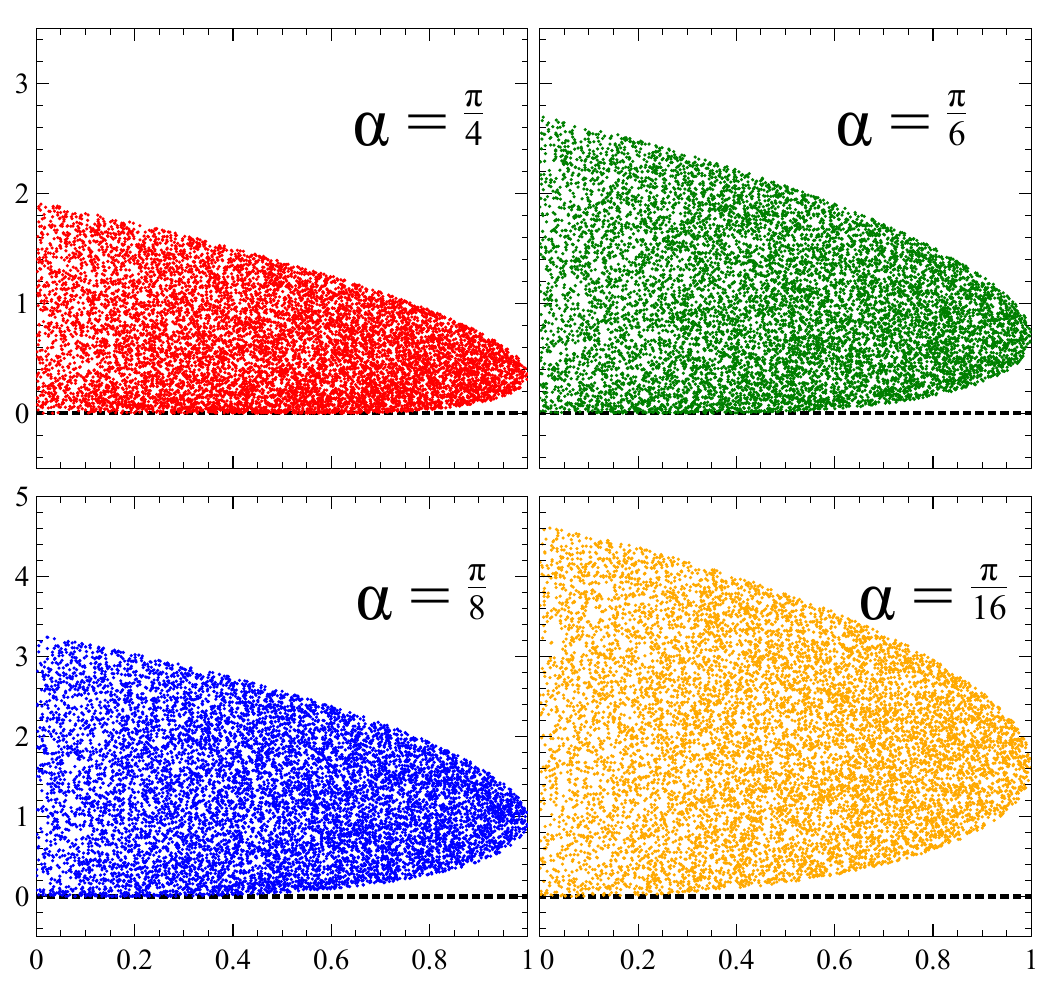}
\caption{(Color online) Scatter diagrams for relative entropy of non-orthogonal coherence (along the vertical axis) vs. von Neumann entropy (along the horizontal axis) for $10^{4}$ Haar uniformly chosen qubit states. 2$\alpha$ is the angle subtended at the origin of the Bloch sphere by the line segment connecting the basis vectors on the Bloch sphere. All axes are measured in bits, and the angles are in radians.}
\label{mcmsrel}
\end{figure}

It is natural to wonder whether we can generalize these notions of NOMCMS and NOMinCMS to arbitrary distance measures in \eqref{cno}. In particular, taking relative entropy as the distance measure, we numerically investigate the connection between the non-orthogonal coherence measure $C_{rel}^{NO}$ and the corresponding  measure of mixedness, which in this case can be quantified by the von Neumann entropy, \(S(\rho)\), of the state $\rho$. 
Specifically, we analyze $C^{NO}_{rel} (\rho)$ for a fixed $S(\rho)$ where $\rho$ is generated Haar uniformly. We again observe the existence of non-trivial upper and lower bounds on the non-orthogonal coherence for a given mixedness. It is clear from Fig.  \ref{mcmsrel}, that the connection between $C^{NO}_{rel} (\rho)$ and $S(\rho)$ depends on the inner product between the basis vectors. We also notice the existence of a purity threshold beyond which there exists non-trivial minimally coherent mixed states. This threshold, as in the case of trace distance-based non-orthogonal coherence measure $C^{NO}_{trace}(\rho)$, is lowered as the angle between the basis vectors decreases.  

\section{ Non-Orthogonal coherence in more than one basis }
\label{anuchhed-chhoi}

In case of orthogonal qubit bases, basis vectors always lie at antipodal points of the Bloch sphere. This constraint is absent for non-orthogonal bases. Thus we can investigate the behavior of non-orthogonal coherence for some possible configurations of non-orthogonal bases which have no analogs for orthogonal bases.  We study below the behavior of quantum coherence in non-orthogonal bases for two such configurations. 

While it is true that there is usually a single natural basis in a given experimental set-up,
quantum mechanics treats all bases with equal ``weightage''. This has led researchers to define coherence 
for a randomly-chosen basis. Motivation for doing so includes the expectation that there can be experimental set-ups 
where there are more than one natural bases. See e.g. \cite{Hall,Yao}. With respect to the experimental set-up 
considered in this paper, different non-orthogonal bases can appear by changing the character of the ``device'' 
just after the slits on the upper beam in Fig. \ref{fig-double-slit}, although it is clear that 
for a given configuration of the said device, we are dealing with a single non-orthogonal basis.

\subsection{Mutually orthogonal bases}
Let us begin with the following scenario. Two non-orthogonal bases are chosen in such a way that the  basis vectors of one basis are perpendicular to the basis vectors of the other. In this case, we call one of the bases as being mutually orthogonal to the other. More concretely, let us choose the basis $K_{1}$ as the set $\lbrace |0\ket, |\psi\ket \rbrace$ and  $ K_{2}$ as  $\lbrace |1\ket, |\psi^{\perp}\ket \rbrace$ with $|\psi\ket$ being represented by angles $(\theta_{0},\phi_{0})$ on the Bloch sphere and $|\psi^{\perp}\ket$ being its antipodal point. Without loss of generality, we can take $ \phi_{0}=0 $.
In this case, the following bound on the sum of non-orthogonal coherences, using the trace distance measure, are obtained.
\\

\emph{\textbf{Proposition IV:} 
The sums of trace distance-based coherences with respect to two mutually orthogonal bases for any qubit state $\rho$ with Bloch sphere representation $(r,\theta, \phi)$ satisfies
\begin{eqnarray}
\label{compl1}
 C_{trace}^{NO(K_{1})} +  C_{trace}^{NO(K_{2})}   \geq  2 \cos \alpha, \hspace{0.4cm}
\end{eqnarray}
where $``K_{1}"$ and $``K_{2}"$ in superscripts are intended to indicate the non-orthogonal basis with respect to which the non-orthogonal coherence is calculated. The relations are valid for all $(r, \theta, \phi)$ and $2 \alpha$ is the angle subtended at the origin by the elements of any of the bases.}

\emph{Proof:}  The trace distance-based coherence of the state of a qubit with respect to a particular non-orthogonal basis, is equal, in the Bloch sphere representation, 
 to the shortest Euclidean distance of that state from the straight line segment joining the two basis vectors constituting that non-orthogonal basis. 
 The limiting values of $ (r, \theta, \phi )$, for which it will be possible to draw a perpendicular on the line segment joining $\lbrace |0\ket, |\psi\ket \rbrace$ and that joining 
  $\lbrace |1\ket, |\psi^{\perp}\ket \rbrace$, are given by 

%
\begin{equation}
\label{A}
r \sin \theta \cos \phi \cos \alpha  \pm  r \cos \theta \sin \alpha \leq  \sin \alpha
\end{equation}
However, irrespective of whether these relations are satisfied, the triangle inequality for the Euclidean norm implies that the sum of the coherences is greater than or equal to the distance between the lines representing the bases $ K_{1} $ an $ K_{2} $. This distance is $ 2 \cos \alpha $, remembering that $ 0 < \alpha < \pi/2 $, in non-trivial cases.

\textbf{Remark.} For states $ \rho $ that fall outside the region defined by inequalities in Eq. (\ref{A}), the coherence, $ C_{trace}^{NO(K_{1})} $ is the shorter of the distances to the states $ |0\rangle $ and $|\psi\rangle $. Similarly, for $ C_{trace}^{NO(K_{2})} $.

\subsection{Cyclic bases}
\begin{figure}
\includegraphics[scale=0.5]{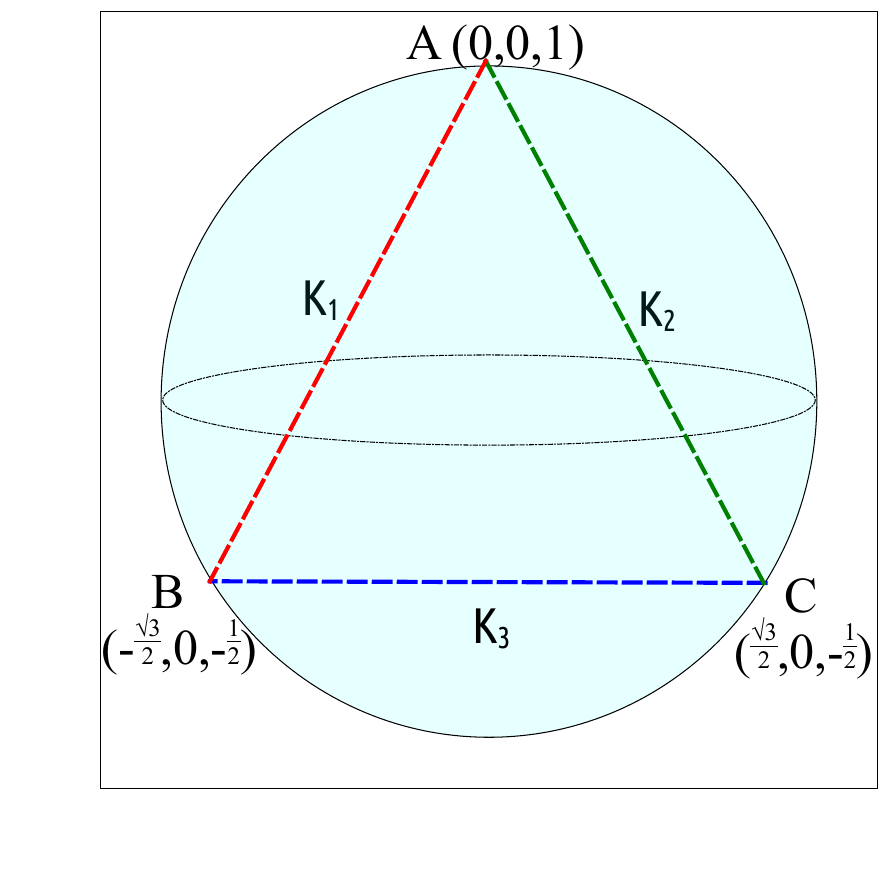}
\caption{(Color online) Basis vectors of $K_{1}, K_{2}, K_{3}$ lying on the great circle along the xz plane. $|AB| = |BC| = |CA|$. }
\label{fig2}
\end{figure}

We now move to a scenario where different non-orthogonal bases, $\lbrace K_{i} \rbrace$, are arranged in such a way that the corresponding basis vectors lie on a great circle and form the vertices of a regular polygon. These bases can be called cyclic bases. If this polygon is an equilateral triangle, as in Fig. \ref{fig2}, then the sum of squared non-orthogonal coherences follows the relation in the proposition below.\\
\begin{figure}
\includegraphics[scale=0.5]{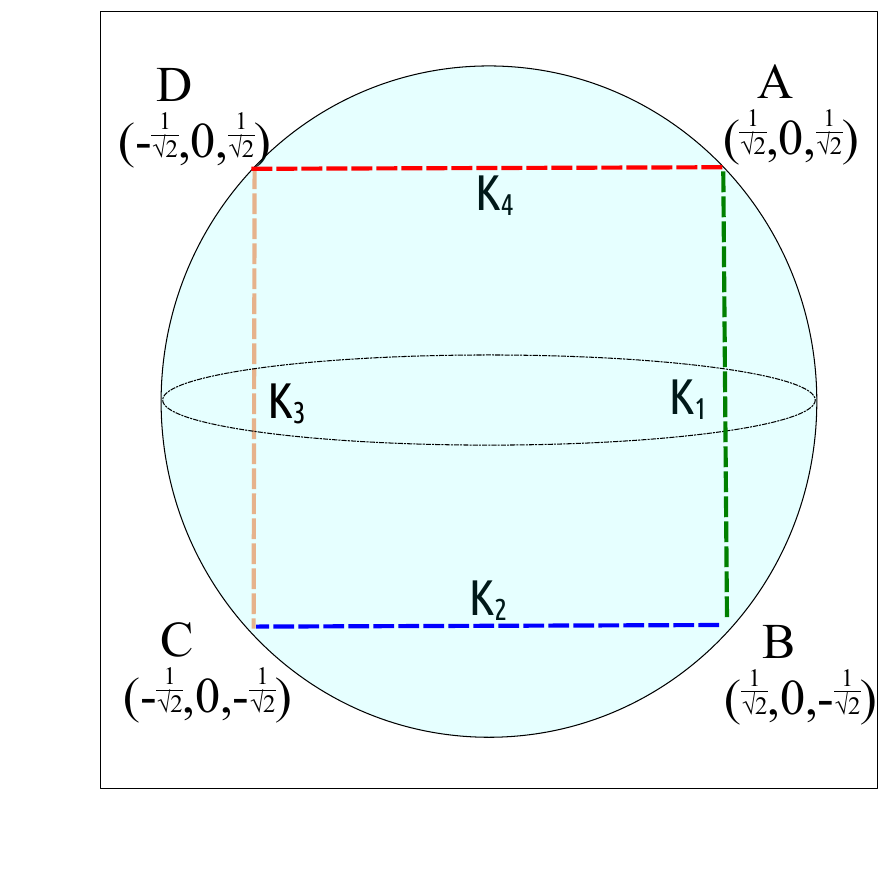}
\caption{(Color online) Basis vectors of $K_{1}, K_{2}, K_{3}, K_{4}$ lying on the great circle along the xz plane. $|AB| = |BC| = |CD| = |DA|$.  } 
\label{fig3}
\end{figure}
\\
\emph{\textbf{Proposition V:} For the cyclic bases $K_{1}, K_{2}, K_{3}$ forming the vertices of an equilateral triangle, the sum of squared trace distance-based coherences for any qubit state with Bloch sphere representation $(r,\theta, \phi)$ satisfies
\small{ \begin{eqnarray}
\label{3gon}
\left(C_{trace}^{NO(K_{1})}\right)^{2}+ \left(C_{trace}^{NO(K_{2})}\right)^{2} +\left(C_{trace}^{NO(K_{3})}\right)^{2} &\geq& \frac{3}{4} (1+ 2 r^{2}) \hspace{0.2 cm} \forall r, \nonumber  \\ 
\left(C_{trace}^{NO(K_{1})}\right)^{2}+ \left(C_{trace}^{NO(K_{2})}\right)^{2} +\left(C_{trace}^{NO(K_{3})}\right)^{2} &\leq& \frac{3}{4} (1+ 4 r^{2})~~\nonumber \\
 \mbox{for} ~ r \leq \frac{1}{2}.
\end{eqnarray}}}

\emph{Proof-}  For any quantum bit represented by the point $(r,\theta.\phi) $ in the Bloch sphere, we find that the sum of squared trace distance-based coherences is exactly equal to the following expression for $r \leq \frac{1}{2}$, and lower bounded by it otherwise:

\beq{
\frac{3}{4} (1+ 4 r^{2}) - \frac{3}{2}r^{2} (\cos^{2}\theta +  \cos^{2}\phi \sin^{2}\theta)  . }\eeq
It can be shown that the minimum   with respect to $\lbrace \theta, \phi \rbrace$ of the above quantity is $\frac{3}{4} (1+ 2 r^{2})$ and the maximum for $r \leq \frac{1}{2}$ is $\frac{3}{4} (1+ 4 r^{2})$. The lower bound may be saturated if we take $\phi= 0$, and the upper bound if we take $\theta = \frac{\pi}{2},\phi = \frac{\pi}{2}$.
 \qed\\

\noindent Let us now consider four non-orthogonal bases lying on a great circle and forming a square as shown in Fig. \ref{fig3}. In this case, we have the following proposition, which can be proved in exactly the same way as before.
\\

\emph{\textbf{Proposition VI:} For the cyclic bases $K_{1}, K_{2}, K_{3}, K_{4}$ forming the vertices of a square, the sum of squared trace distance-based coherences for any qubit state with Bloch sphere representation $(r,\theta, \phi)$  satisfies
\begin{eqnarray}
\label{4gon}
\sum_{i= 1}^{4} \left(C_{trace}^{NO(K_{i})}\right)^{2}  \geq 2 (1+  r^{2}) \hspace{0.2 cm} \forall r, \nonumber  \\ 
\sum_{i= 1}^{4} \left(C_{trace}^{NO(K_{i})}\right)^{2}  \leq 2 (1+ 2 r^{2})~~ \mbox{for} ~ r \leq \frac{1}{\sqrt{2}}.
\end{eqnarray}}

For a given set of bases, we can define the ``total non-orthogonal coherence'' as the sum over the 
non-orthogonal coherences in those bases.   
From the relations in propositions V and VI, we note that for any fixed purity $r$,  and for a given set of bases, along with having a state with the maximum amount of total non-orthogonal coherence,  there also exists  states with a non-trivial minimum amount of total non-orthogonal coherence. While the  former may be seen as a generalization of MCMS to the case of non-orthogonal bases, the latter is interesting in its own right and the  states saturating the lower bounds in propositions V and VI may be identified as generalizations of non-orthogonal minimally coherent mixed states discussed in the previous section, although in this case, there is no purity threshold $\tilde{r}$, which was required for the existence of $NOMinCMS$. We note that the states 
with minimum and maximum total non-orthogonal coherence do actually exist.

Generalizing to a regular polygon having N different bases, $N \rightarrow \infty$  
corresponds to the situation where each point on a certain great circle is a non-orthogonal basis (in the limiting sense).
It may be verified that every point on the  plane corresponding to this great circle has a total non-orthogonal coherence that depends only on the purity.


\begin{figure}
\includegraphics[scale=0.5]{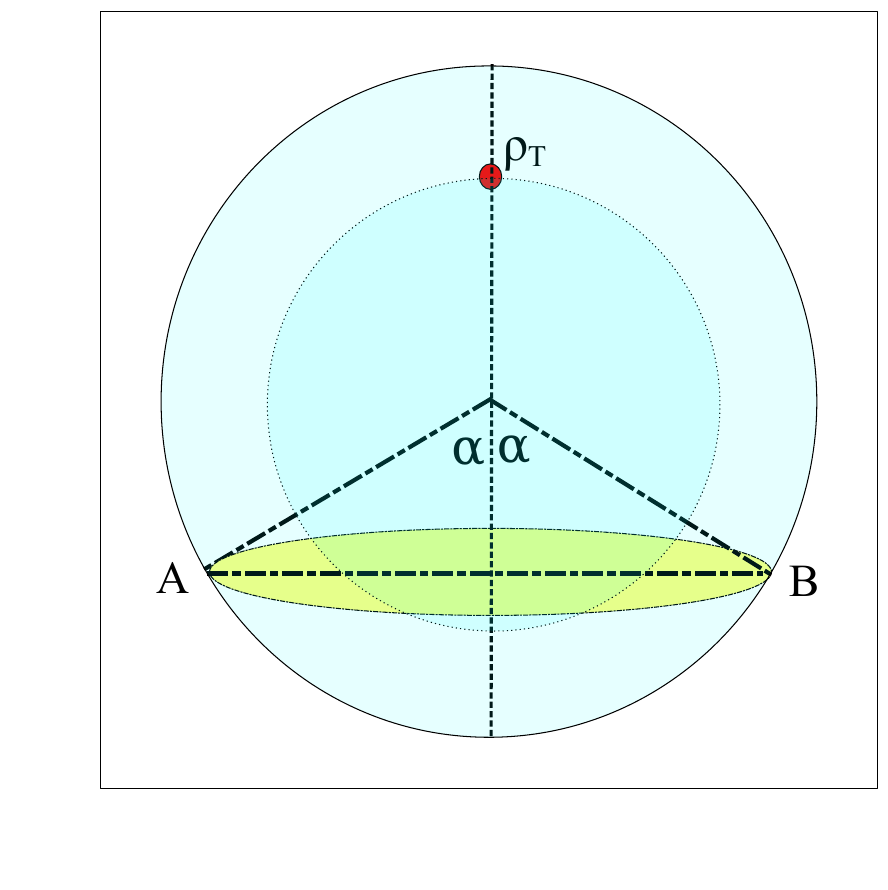}
\caption{(Color online) $\rho^{T}$ is the thermal state incoherent in the energy eigenbasis formed by the extremities of the dotted line. AB is one of the bases in which $\rho^{T}$ is a non-orthogonal MCMS. $\alpha$ is the inner product angle between the basis vectors of this basis. $\rho^{T}$ is a MCMS for any basis formed by vectors at two radially opposite points of the green disc.}
\label{fig4}
\end{figure}

\section{Energy Cost of Basis-Changing Operations and its Relation to Non-Orthogonal Coherence}
\label{anuchhed-saath}

In this section, we find the energy cost of a basis-changing operation from an energy 
eigenbasis to a non-orthogonal basis. The non-orthogonal basis is chosen so that the thermal state (whose eigenbasis is just the energy eigenbasis) is maximally coherent in that basis, among all states of the same purity as that of the thermal state. We then find the relation of this energy cost with the non-orthogonal coherence of the thermal state (before the basis-changing operation).

Suppose we have a qubit thermal state $\rho^{T} = \frac{e^{-\beta H}}{Z}$, diagonal in the computational basis. The computational basis is taken here as the energy eigenbasis. Here, $\beta = \frac{1}{k_BT}$, $ T $  is the absolute temperature, and $k_{B}$ is the Boltzmann constant.  
This thermal state is maximally coherent for any non-orthogonal basis $\lbrace|b_{1}\ket = \cos\frac{\pi-\alpha}{2} |0\ket + e^{i\phi} \sin \frac{\pi-\alpha}{2} |1\ket, |b_{2}\ket = \cos\frac{\pi-\alpha}{2} |0\ket - e^{i\phi} \sin \frac{\pi-\alpha}{2} |1\ket \rbrace$, with $0 < \alpha \leq \frac{\pi}{2}$, while it is incoherent with respect to the energy eigenbasis. We wish to quantify the energy cost of the change in the basis from the orthonormal computational basis to the non-orthogonal basis.  This cost can be interpreted as that required to correspondingly transform the apparatuses in a possible experimental implementation of the relevant BC operation. As discussed before, 
transforming an orthogonal basis into a non-orthogonal one in a system (S) necessarily requires an auxiliary system (A). If we denote the corresponding unitary by \(U_{SA}\), its action 
is given by 
\begin{eqnarray}
\label{eita-energy}
|0\ket_{S} |b_{1}\ket_{A} \longrightarrow |b_{1}\ket_{S} |0\ket_{A}, \nonumber \\
|1\ket_{S} |b{1}\ket_{A} \longrightarrow |b_{2}\ket_{S} |1\ket_{A}.
\end{eqnarray}
A possible mechanism to implement this unitary when we are allowed to perform only controlled unitary operations and the swap operation, is as follows. 
(It is easy to see that the phase \(\phi\) has no consequence in the considerations below, and is therefore dropped.)

\begin{enumerate}

\item \emph{Controlled unitary -} We apply a controlled unitary gate, $U^{CU}$, on the system and auxiliary which keeps the joint state $|0\ket_{S} \otimes |b_{1} \ket_{A} $  invariant and transforms the joint state $|1\ket_{S} \otimes |b_{1} \ket_{A} $ into $|1\ket_{S} \otimes U |b_{1} \ket_{A} = |1\ket_{S} \otimes  |b_{2} \ket_{A} $ where \(U\) is a single qubit unitary which takes $|b_{1}\ket$ into $|b_{2}\ket$. 

\item \emph{Swap -} We swap the system and the auxiliary qubits. 


\end{enumerate}

\noindent The total effect of these steps is the application of the global unitary  $U_{SA}$.

We may now identify the energy cost of the change in basis with the change in energy of SA, and it is given by  
\small{
\begin{eqnarray}
\Delta = \tr  \left[ \left(H_{S} \otimes \mathbb{I}_{A} \right)\left( \rho^{T}_{S}\otimes |b_{1}\ket_{A}\bra b_{1}| - U_{SA} (\rho^{T}_{S} \otimes |b_{1}\ket_{A}\bra b_{1}|) U_{SA}^{\dagger}   \right) \right] \nonumber,
\end{eqnarray}}
where $ H_{S} $ is the Hamiltonian of the system, with its two energy levels being $ |0\rangle $ and $ |1\rangle $. We have assumed here that the auxiliary system is initially in the state $ |b_{1} \rangle$.
The trace distance-based non-orthogonal coherence in the $ \{ |b_{1}\rangle, |b_{2}\rangle \} $ basis of the  thermal state is given by 
\begin{eqnarray}
C^{NO}_{trace} = r(\rho^{T}) + \cos \alpha ,
\end{eqnarray}
\noindent with $r$, i.e. the Bloch radius, being a measure of purity of the thermal state, \(\rho^T\). This non-orthogonal coherence may be interpreted as the coherence generated by the unitary operator $ U_{SA} $.

Explicitly calculating the energy cost with the system Hamiltonian, \(H_S = E_{1} |1\rangle\langle 1|\), being that of a two level system, with energy levels $0$ and $E_{1} > 0$, yields

\begin{equation}
\label{energycost}
\Delta =- \frac{1}{2} E_{1} \left(\cos \alpha + \tanh \frac{E_{1}}{2 k_B T} \right).
\end{equation} 

\noindent Putting in the expression of non-orthogonal trace distance-based coherence $C_{trace}^{NO}$, we obtain the relation
\begin{equation}
\label{energycost2} 
\Delta = -\frac{1}{2} E_{1} C_{trace}^{NO}.
\end{equation}
The relation \eqref{energycost2} reveals that the energy cost of non-orthogonal coherence is linearly proportional to the non-orthogonal coherence created.
 
\vspace{0.3cm}

\section{Conclusion}
\label{anuchhed-aat}

In summary, we have proposed a resource theory of
coherence, based on a fixed non-orthogonal basis, and have introduced
corresponding states and operations which do not incur a cost in this
theory. The free operations in this theory are connected with the same in
the resource theory of coherence with orthonormal bases. We also find a
linear relation between the energy cost of generating non-orthogonal
coherence and non-orthogonal coherence created. There are several works in the literature that are similar to the one presented in this paper. Notable ones include Refs. \cite{plenio2,Rastegin,Bischof1,Bischof2}. Their differences with the present one have been mentioned in the text.

There are several aspects of the resource theory of non-orthogonal quantum
coherence that are devoid of analogs in the case of orthonormal bases. In
particular, there is a unique maximally coherent qubit state for a given
non-orthogonal basis. The resource theory of coherence for orthonormal
bases allows for an infinite number of maximally resourceful states. 
Perhaps more strikingly, there is a minimal non-orthogonal
coherence that all qubit states must have for any non-orthogonal basis, provided their 
purity is sufficiently low.
This naturally led us to obtain tight bounds on the sum and difference of
non-orthogonal coherence and purity for arbitrary qubit states. For
multiple families of non-orthogonal bases - families without counterparts
in the scenario of orthonormal bases - we find strong constraints on the
sum of squares of the non-orthogonal coherences corresponding to the bases
belonging to the families.

We believe that the results obtained will be important for fundamental as
well as applicational aspects of quantum information. In particular, we uncover a 
wave-particle-type duality as seen in a complementarity
 between the non-orthogonal quantum coherence and which-path information in 
a quantum double-slit set-up with a leak. The same relation also holds in 
 a leaky Mach-Zehnder interferometric set-up.
 
\acknowledgements
This research was supported in part by the INFOSYS scholarship for senior
students.

\bibliography{noc}

\begin{thebibliography}{90}%
\makeatletter
\providecommand \@ifxundefined [1]{%
 \@ifx{#1\undefined}
}%
\providecommand \@ifnum [1]{%
 \ifnum #1\expandafter \@firstoftwo
 \else \expandafter \@secondoftwo
 \fi
}%
\providecommand \@ifx [1]{%
 \ifx #1\expandafter \@firstoftwo
 \else \expandafter \@secondoftwo
 \fi
}%
\providecommand \natexlab [1]{#1}%
\providecommand \enquote  [1]{``#1''}%
\providecommand \bibnamefont  [1]{#1}%
\providecommand \bibfnamefont [1]{#1}%
\providecommand \citenamefont [1]{#1}%
\providecommand \href@noop [0]{\@secondoftwo}%
\providecommand \href [0]{\begingroup \@sanitize@url \@href}%
\providecommand \@href[1]{\@@startlink{#1}\@@href}%
\providecommand \@@href[1]{\endgroup#1\@@endlink}%
\providecommand \@sanitize@url [0]{\catcode `\\12\catcode `\$12\catcode
  `\&12\catcode `\#12\catcode `\^12\catcode `\_12\catcode `\%12\relax}%
\providecommand \@@startlink[1]{}%
\providecommand \@@endlink[0]{}%
\providecommand \url  [0]{\begingroup\@sanitize@url \@url }%
\providecommand \@url [1]{\endgroup\@href {#1}{\urlprefix }}%
\providecommand \urlprefix  [0]{URL }%
\providecommand \Eprint [0]{\href }%
\providecommand \doibase [0]{http://dx.doi.org/}%
\providecommand \selectlanguage [0]{\@gobble}%
\providecommand \bibinfo  [0]{\@secondoftwo}%
\providecommand \bibfield  [0]{\@secondoftwo}%
\providecommand \translation [1]{[#1]}%
\providecommand \BibitemOpen [0]{}%
\providecommand \bibitemStop [0]{}%
\providecommand \bibitemNoStop [0]{.\EOS\space}%
\providecommand \EOS [0]{\spacefactor3000\relax}%
\providecommand \BibitemShut  [1]{\csname bibitem#1\endcsname}%
\let\auto@bib@innerbib\@empty
\bibitem [{\citenamefont {Schr{\"o}dinger}(1935)}]{cat}%
  \BibitemOpen
  \bibfield  {author} {\bibinfo {author} {\bibfnamefont {E.}~\bibnamefont
  {Schr{\"o}dinger}},\ }\href {\doibase 10.1007/BF01491891} {\bibfield
  {journal} {\bibinfo  {journal} {Naturwissenschaften}\ }\textbf {\bibinfo
  {volume} {23}},\ \bibinfo {pages} {807} (\bibinfo {year} {1935})}\BibitemShut
  {NoStop}%
\bibitem [{\citenamefont {Wootters}\ and\ \citenamefont
  {Zurek}(1982)}]{nocloning}%
  \BibitemOpen
  \bibfield  {author} {\bibinfo {author} {\bibfnamefont {W.~K.}\ \bibnamefont
  {Wootters}}\ and\ \bibinfo {author} {\bibfnamefont {W.~H.}\ \bibnamefont
  {Zurek}},\ }\href {\doibase 10.1038/299802a0} {\bibfield  {journal} {\bibinfo
   {journal} {Nature}\ }\textbf {\bibinfo {volume} {299}},\ \bibinfo {pages}
  {802} (\bibinfo {year} {1982})}\BibitemShut {NoStop}%
\bibitem [{\citenamefont {{Pati}}\ and\ \citenamefont
  {{Braunstein}}(2000)}]{nodeleting}%
  \BibitemOpen
  \bibfield  {author} {\bibinfo {author} {\bibfnamefont {A.~K.}\ \bibnamefont
  {{Pati}}}\ and\ \bibinfo {author} {\bibfnamefont {S.~L.}\ \bibnamefont
  {{Braunstein}}},\ }\href@noop {} {\bibfield  {journal} {\bibinfo  {journal}
  {\nat}\ }\textbf {\bibinfo {volume} {404}},\ \bibinfo {pages} {164} (\bibinfo
  {year} {2000})}\BibitemShut {NoStop}%
\bibitem [{\citenamefont {Barnum}\ \emph {et~al.}(1996)\citenamefont {Barnum},
  \citenamefont {Caves}, \citenamefont {Fuchs}, \citenamefont {Jozsa},\ and\
  \citenamefont {Schumacher}}]{barnum}%
  \BibitemOpen
  \bibfield  {author} {\bibinfo {author} {\bibfnamefont {H.}~\bibnamefont
  {Barnum}}, \bibinfo {author} {\bibfnamefont {C.~M.}\ \bibnamefont {Caves}},
  \bibinfo {author} {\bibfnamefont {C.~A.}\ \bibnamefont {Fuchs}}, \bibinfo
  {author} {\bibfnamefont {R.}~\bibnamefont {Jozsa}}, \ and\ \bibinfo {author}
  {\bibfnamefont {B.}~\bibnamefont {Schumacher}},\ }\href {\doibase
  10.1103/PhysRevLett.76.2818} {\bibfield  {journal} {\bibinfo  {journal}
  {Phys. Rev. Lett.}\ }\textbf {\bibinfo {volume} {76}},\ \bibinfo {pages}
  {2818} (\bibinfo {year} {1996})}\BibitemShut {NoStop}%
\bibitem [{\citenamefont {Bennett}\ and\ \citenamefont
  {Wiesner}(1992)}]{wiesner}%
  \BibitemOpen
  \bibfield  {author} {\bibinfo {author} {\bibfnamefont {C.~H.}\ \bibnamefont
  {Bennett}}\ and\ \bibinfo {author} {\bibfnamefont {S.~J.}\ \bibnamefont
  {Wiesner}},\ }\href {\doibase 10.1103/PhysRevLett.69.2881} {\bibfield
  {journal} {\bibinfo  {journal} {Phys. Rev. Lett.}\ }\textbf {\bibinfo
  {volume} {69}},\ \bibinfo {pages} {2881} (\bibinfo {year}
  {1992})}\BibitemShut {NoStop}%
\bibitem [{\citenamefont {Bennett}\ \emph {et~al.}(1993)\citenamefont
  {Bennett}, \citenamefont {Brassard}, \citenamefont {Cr\'epeau}, \citenamefont
  {Jozsa}, \citenamefont {Peres},\ and\ \citenamefont
  {Wootters}}]{teleportation}%
  \BibitemOpen
  \bibfield  {author} {\bibinfo {author} {\bibfnamefont {C.~H.}\ \bibnamefont
  {Bennett}}, \bibinfo {author} {\bibfnamefont {G.}~\bibnamefont {Brassard}},
  \bibinfo {author} {\bibfnamefont {C.}~\bibnamefont {Cr\'epeau}}, \bibinfo
  {author} {\bibfnamefont {R.}~\bibnamefont {Jozsa}}, \bibinfo {author}
  {\bibfnamefont {A.}~\bibnamefont {Peres}}, \ and\ \bibinfo {author}
  {\bibfnamefont {W.~K.}\ \bibnamefont {Wootters}},\ }\href {\doibase
  10.1103/PhysRevLett.70.1895} {\bibfield  {journal} {\bibinfo  {journal}
  {Phys. Rev. Lett.}\ }\textbf {\bibinfo {volume} {70}},\ \bibinfo {pages}
  {1895} (\bibinfo {year} {1993})}\BibitemShut {NoStop}%
\bibitem [{\citenamefont {{Shor}}(1995)}]{shor}%
  \BibitemOpen
  \bibfield  {author} {\bibinfo {author} {\bibfnamefont {P.~W.}\ \bibnamefont
  {{Shor}}},\ }\href@noop {} {\bibfield  {journal} {\bibinfo  {journal} {eprint
  arXiv:quant-ph/9508027}\ } (\bibinfo {year} {1995})},\ \Eprint
  {http://arxiv.org/abs/quant-ph/9508027} {quant-ph/9508027} \BibitemShut
  {NoStop}%
\bibitem [{\citenamefont {Pati}(2000)}]{rsp2}%
  \BibitemOpen
  \bibfield  {author} {\bibinfo {author} {\bibfnamefont {A.~K.}\ \bibnamefont
  {Pati}},\ }\href {\doibase 10.1103/PhysRevA.63.014302} {\bibfield  {journal}
  {\bibinfo  {journal} {Phys. Rev. A}\ }\textbf {\bibinfo {volume} {63}},\
  \bibinfo {pages} {014302} (\bibinfo {year} {2000})}\BibitemShut {NoStop}%
\bibitem [{\citenamefont {Bennett}\ \emph {et~al.}(2001)\citenamefont
  {Bennett}, \citenamefont {DiVincenzo}, \citenamefont {Shor}, \citenamefont
  {Smolin}, \citenamefont {Terhal},\ and\ \citenamefont {Wootters}}]{rsp1}%
  \BibitemOpen
  \bibfield  {author} {\bibinfo {author} {\bibfnamefont {C.~H.}\ \bibnamefont
  {Bennett}}, \bibinfo {author} {\bibfnamefont {D.~P.}\ \bibnamefont
  {DiVincenzo}}, \bibinfo {author} {\bibfnamefont {P.~W.}\ \bibnamefont
  {Shor}}, \bibinfo {author} {\bibfnamefont {J.~A.}\ \bibnamefont {Smolin}},
  \bibinfo {author} {\bibfnamefont {B.~M.}\ \bibnamefont {Terhal}}, \ and\
  \bibinfo {author} {\bibfnamefont {W.~K.}\ \bibnamefont {Wootters}},\ }\href
  {\doibase 10.1103/PhysRevLett.87.077902} {\bibfield  {journal} {\bibinfo
  {journal} {Phys. Rev. Lett.}\ }\textbf {\bibinfo {volume} {87}},\ \bibinfo
  {pages} {077902} (\bibinfo {year} {2001})}\BibitemShut {NoStop}%
\bibitem [{\citenamefont {Bennett}\ and\ \citenamefont
  {Brassard}(1984)}]{BB84}%
  \BibitemOpen
  \bibfield  {author} {\bibinfo {author} {\bibfnamefont {C.~H.}\ \bibnamefont
  {Bennett}}\ and\ \bibinfo {author} {\bibfnamefont {G.}~\bibnamefont
  {Brassard}},\ }\href@noop {} {\bibfield  {journal} {\bibinfo  {journal}
  {Proceedings of IEEE International Conference on Computers, Systems and
  Signal Processing}\ }\textbf {\bibinfo {volume} {175}},\ \bibinfo {pages} {8}
  (\bibinfo {year} {1984})}\BibitemShut {NoStop}%
\bibitem [{\citenamefont {Ekert}(1991)}]{e91}%
  \BibitemOpen
  \bibfield  {author} {\bibinfo {author} {\bibfnamefont {A.~K.}\ \bibnamefont
  {Ekert}},\ }\href {\doibase 10.1103/PhysRevLett.67.661} {\bibfield  {journal}
  {\bibinfo  {journal} {Phys. Rev. Lett.}\ }\textbf {\bibinfo {volume} {67}},\
  \bibinfo {pages} {661} (\bibinfo {year} {1991})}\BibitemShut {NoStop}%
\bibitem [{\citenamefont {Bennett}(1992)}]{b92}%
  \BibitemOpen
  \bibfield  {author} {\bibinfo {author} {\bibfnamefont {C.~H.}\ \bibnamefont
  {Bennett}},\ }\href {\doibase 10.1103/PhysRevLett.68.3121} {\bibfield
  {journal} {\bibinfo  {journal} {Phys. Rev. Lett.}\ }\textbf {\bibinfo
  {volume} {68}},\ \bibinfo {pages} {3121} (\bibinfo {year}
  {1992})}\BibitemShut {NoStop}%
\bibitem [{\citenamefont {{Aberg}}(2006)}]{aberg}%
  \BibitemOpen
  \bibfield  {author} {\bibinfo {author} {\bibfnamefont {J.}~\bibnamefont
  {{Aberg}}},\ }\href@noop {} {\bibfield  {journal} {\bibinfo  {journal}
  {eprint arXiv:quant-ph/0612146}\ } (\bibinfo {year} {2006})},\ \Eprint
  {http://arxiv.org/abs/quant-ph/0612146} {quant-ph/0612146} \BibitemShut
  {NoStop}%
\bibitem [{\citenamefont {Baumgratz}\ \emph {et~al.}(2014)\citenamefont
  {Baumgratz}, \citenamefont {Cramer},\ and\ \citenamefont
  {Plenio}}]{baumgratz}%
  \BibitemOpen
  \bibfield  {author} {\bibinfo {author} {\bibfnamefont {T.}~\bibnamefont
  {Baumgratz}}, \bibinfo {author} {\bibfnamefont {M.}~\bibnamefont {Cramer}}, \
  and\ \bibinfo {author} {\bibfnamefont {M.~B.}\ \bibnamefont {Plenio}},\
  }\href {\doibase 10.1103/PhysRevLett.113.140401} {\bibfield  {journal}
  {\bibinfo  {journal} {Phys. Rev. Lett.}\ }\textbf {\bibinfo {volume} {113}},\
  \bibinfo {pages} {140401} (\bibinfo {year} {2014})}\BibitemShut {NoStop}%
\bibitem [{\citenamefont {Winter}\ and\ \citenamefont {Yang}(2016)}]{winter}%
  \BibitemOpen
  \bibfield  {author} {\bibinfo {author} {\bibfnamefont {A.}~\bibnamefont
  {Winter}}\ and\ \bibinfo {author} {\bibfnamefont {D.}~\bibnamefont {Yang}},\
  }\href {\doibase 10.1103/PhysRevLett.116.120404} {\bibfield  {journal}
  {\bibinfo  {journal} {Phys. Rev. Lett.}\ }\textbf {\bibinfo {volume} {116}},\
  \bibinfo {pages} {120404} (\bibinfo {year} {2016})}\BibitemShut {NoStop}%
\bibitem [{\citenamefont {Mondal}\ \emph {et~al.}(2017)\citenamefont {Mondal},
  \citenamefont {Pramanik},\ and\ \citenamefont {Pati}}]{debasis}%
  \BibitemOpen
  \bibfield  {author} {\bibinfo {author} {\bibfnamefont {D.}~\bibnamefont
  {Mondal}}, \bibinfo {author} {\bibfnamefont {T.}~\bibnamefont {Pramanik}}, \
  and\ \bibinfo {author} {\bibfnamefont {A.~K.}\ \bibnamefont {Pati}},\ }\href
  {\doibase 10.1103/PhysRevA.95.010301} {\bibfield  {journal} {\bibinfo
  {journal} {Phys. Rev. A}\ }\textbf {\bibinfo {volume} {95}},\ \bibinfo
  {pages} {010301} (\bibinfo {year} {2017})}\BibitemShut {NoStop}%
\bibitem [{\citenamefont {{Bu}}\ \emph {et~al.}(2016)\citenamefont {{Bu}},
  \citenamefont {{Kumar}},\ and\ \citenamefont {{Wu}}}]{nloc1}%
  \BibitemOpen
  \bibfield  {author} {\bibinfo {author} {\bibfnamefont {K.}~\bibnamefont
  {{Bu}}}, \bibinfo {author} {\bibfnamefont {A.}~\bibnamefont {{Kumar}}}, \
  and\ \bibinfo {author} {\bibfnamefont {J.}~\bibnamefont {{Wu}}},\ }\href@noop
  {} {\bibfield  {journal} {\bibinfo  {journal} {ArXiv e-prints}\ } (\bibinfo
  {year} {2016})},\ \Eprint {http://arxiv.org/abs/1603.06322} {arXiv:1603.06322
  [quant-ph]} \BibitemShut {NoStop}%
\bibitem [{\citenamefont {{Qiu}}\ \emph {et~al.}(2016)\citenamefont {{Qiu}},
  \citenamefont {{Liu}},\ and\ \citenamefont {{Pan}}}]{nloc2}%
  \BibitemOpen
  \bibfield  {author} {\bibinfo {author} {\bibfnamefont {L.}~\bibnamefont
  {{Qiu}}}, \bibinfo {author} {\bibfnamefont {Z.}~\bibnamefont {{Liu}}}, \ and\
  \bibinfo {author} {\bibfnamefont {F.}~\bibnamefont {{Pan}}},\ }\href@noop {}
  {\bibfield  {journal} {\bibinfo  {journal} {ArXiv e-prints}\ } (\bibinfo
  {year} {2016})},\ \Eprint {http://arxiv.org/abs/1610.07237} {arXiv:1610.07237
  [quant-ph]} \BibitemShut {NoStop}%
\bibitem [{\citenamefont {{Chanda}}\ and\ \citenamefont
  {{Bhattacharya}}(2016)}]{titas}%
  \BibitemOpen
  \bibfield  {author} {\bibinfo {author} {\bibfnamefont {T.}~\bibnamefont
  {{Chanda}}}\ and\ \bibinfo {author} {\bibfnamefont {S.}~\bibnamefont
  {{Bhattacharya}}},\ }\href {\doibase 10.1016/j.aop.2016.01.004} {\bibfield
  {journal} {\bibinfo  {journal} {Annals of Physics}\ }\textbf {\bibinfo
  {volume} {366}},\ \bibinfo {pages} {1} (\bibinfo {year} {2016})}\BibitemShut
  {NoStop}%
\bibitem [{\citenamefont {{Bhattacharya}}\ \emph {et~al.}(2016)\citenamefont
  {{Bhattacharya}}, \citenamefont {{Banerjee}},\ and\ \citenamefont
  {{Pati}}}]{nm1}%
  \BibitemOpen
  \bibfield  {author} {\bibinfo {author} {\bibfnamefont {S.}~\bibnamefont
  {{Bhattacharya}}}, \bibinfo {author} {\bibfnamefont {S.}~\bibnamefont
  {{Banerjee}}}, \ and\ \bibinfo {author} {\bibfnamefont {A.~K.}\ \bibnamefont
  {{Pati}}},\ }\href@noop {} {\bibfield  {journal} {\bibinfo  {journal} {ArXiv
  e-prints}\ } (\bibinfo {year} {2016})},\ \Eprint
  {http://arxiv.org/abs/1601.04742} {arXiv:1601.04742 [quant-ph]} \BibitemShut
  {NoStop}%
\bibitem [{\citenamefont {Huang}\ and\ \citenamefont {Situ}(2017)}]{nm2}%
  \BibitemOpen
  \bibfield  {author} {\bibinfo {author} {\bibfnamefont {Z.}~\bibnamefont
  {Huang}}\ and\ \bibinfo {author} {\bibfnamefont {H.}~\bibnamefont {Situ}},\
  }\href {\doibase 10.1007/s10773-016-3192-7} {\bibfield  {journal} {\bibinfo
  {journal} {International Journal of Theoretical Physics}\ }\textbf {\bibinfo
  {volume} {56}},\ \bibinfo {pages} {503} (\bibinfo {year} {2017})}\BibitemShut
  {NoStop}%
\bibitem [{\citenamefont {{{\c C}akmak}}\ \emph {et~al.}(2017)\citenamefont
  {{{\c C}akmak}}, \citenamefont {{Pezzutto}}, \citenamefont {{Paternostro}},\
  and\ \citenamefont {{M{\"u}stecapl{\i}o{\u g}lu}}}]{nm3}%
  \BibitemOpen
  \bibfield  {author} {\bibinfo {author} {\bibfnamefont {B.}~\bibnamefont {{{\c
  C}akmak}}}, \bibinfo {author} {\bibfnamefont {M.}~\bibnamefont {{Pezzutto}}},
  \bibinfo {author} {\bibfnamefont {M.}~\bibnamefont {{Paternostro}}}, \ and\
  \bibinfo {author} {\bibfnamefont {{\"O}.~E.}\ \bibnamefont
  {{M{\"u}stecapl{\i}o{\u g}lu}}},\ }\href@noop {} {\bibfield  {journal}
  {\bibinfo  {journal} {ArXiv e-prints}\ } (\bibinfo {year} {2017})},\ \Eprint
  {http://arxiv.org/abs/1702.05323} {arXiv:1702.05323 [quant-ph]} \BibitemShut
  {NoStop}%
\bibitem [{\citenamefont {{Mukhopadhyay}}\ \emph {et~al.}(2017)\citenamefont
  {{Mukhopadhyay}}, \citenamefont {{Bhattacharya}}, \citenamefont {{Misra}},\
  and\ \citenamefont {{Pati}}}]{nm4}%
  \BibitemOpen
  \bibfield  {author} {\bibinfo {author} {\bibfnamefont {C.}~\bibnamefont
  {{Mukhopadhyay}}}, \bibinfo {author} {\bibfnamefont {S.}~\bibnamefont
  {{Bhattacharya}}}, \bibinfo {author} {\bibfnamefont {A.}~\bibnamefont
  {{Misra}}}, \ and\ \bibinfo {author} {\bibfnamefont {A.~K.}\ \bibnamefont
  {{Pati}}},\ }\href@noop {} {\bibfield  {journal} {\bibinfo  {journal} {ArXiv
  e-prints}\ } (\bibinfo {year} {2017})},\ \Eprint
  {http://arxiv.org/abs/1704.08291} {arXiv:1704.08291 [quant-ph]} \BibitemShut
  {NoStop}%
\bibitem [{\citenamefont {{Mortezapour}}\ \emph {et~al.}(2017)\citenamefont
  {{Mortezapour}}, \citenamefont {{Ahmadi Borji}},\ and\ \citenamefont {{Lo
  Franco}}}]{nm5}%
  \BibitemOpen
  \bibfield  {author} {\bibinfo {author} {\bibfnamefont {A.}~\bibnamefont
  {{Mortezapour}}}, \bibinfo {author} {\bibfnamefont {M.}~\bibnamefont {{Ahmadi
  Borji}}}, \ and\ \bibinfo {author} {\bibfnamefont {R.}~\bibnamefont {{Lo
  Franco}}},\ }\href@noop {} {\bibfield  {journal} {\bibinfo  {journal} {ArXiv
  e-prints}\ } (\bibinfo {year} {2017})},\ \Eprint
  {http://arxiv.org/abs/1705.00887} {arXiv:1705.00887 [quant-ph]} \BibitemShut
  {NoStop}%
\bibitem [{\citenamefont {Streltsov}\ \emph {et~al.}(2015)\citenamefont
  {Streltsov}, \citenamefont {Singh}, \citenamefont {Dhar}, \citenamefont
  {Bera},\ and\ \citenamefont {Adesso}}]{uttament}%
  \BibitemOpen
  \bibfield  {author} {\bibinfo {author} {\bibfnamefont {A.}~\bibnamefont
  {Streltsov}}, \bibinfo {author} {\bibfnamefont {U.}~\bibnamefont {Singh}},
  \bibinfo {author} {\bibfnamefont {H.~S.}\ \bibnamefont {Dhar}}, \bibinfo
  {author} {\bibfnamefont {M.~N.}\ \bibnamefont {Bera}}, \ and\ \bibinfo
  {author} {\bibfnamefont {G.}~\bibnamefont {Adesso}},\ }\href {\doibase
  10.1103/PhysRevLett.115.020403} {\bibfield  {journal} {\bibinfo  {journal}
  {Phys. Rev. Lett.}\ }\textbf {\bibinfo {volume} {115}},\ \bibinfo {pages}
  {020403} (\bibinfo {year} {2015})}\BibitemShut {NoStop}%
\bibitem [{\citenamefont {Asb\'oth}\ \emph {et~al.}(2005)\citenamefont
  {Asb\'oth}, \citenamefont {Calsamiglia},\ and\ \citenamefont
  {Ritsch}}]{ent1}%
  \BibitemOpen
  \bibfield  {author} {\bibinfo {author} {\bibfnamefont {J.~K.}\ \bibnamefont
  {Asb\'oth}}, \bibinfo {author} {\bibfnamefont {J.}~\bibnamefont
  {Calsamiglia}}, \ and\ \bibinfo {author} {\bibfnamefont {H.}~\bibnamefont
  {Ritsch}},\ }\href {\doibase 10.1103/PhysRevLett.94.173602} {\bibfield
  {journal} {\bibinfo  {journal} {Phys. Rev. Lett.}\ }\textbf {\bibinfo
  {volume} {94}},\ \bibinfo {pages} {173602} (\bibinfo {year}
  {2005})}\BibitemShut {NoStop}%
\bibitem [{\citenamefont {Xi}\ \emph {et~al.}(2015)\citenamefont {Xi},
  \citenamefont {Li},\ and\ \citenamefont {Fan}}]{ent2}%
  \BibitemOpen
  \bibfield  {author} {\bibinfo {author} {\bibfnamefont {Z.}~\bibnamefont
  {Xi}}, \bibinfo {author} {\bibfnamefont {Y.}~\bibnamefont {Li}}, \ and\
  \bibinfo {author} {\bibfnamefont {H.}~\bibnamefont {Fan}},\ }\href
  {https://www.nature.com/articles/srep10922} {\bibfield  {journal} {\bibinfo
  {journal} {Sci. Rep.}\ }\textbf {\bibinfo {volume} {5}},\ \bibinfo {pages}
  {10922} (\bibinfo {year} {2015})}\BibitemShut {NoStop}%
\bibitem [{\citenamefont {Streltsov}\ \emph {et~al.}(2016)\citenamefont
  {Streltsov}, \citenamefont {Chitambar}, \citenamefont {Rana}, \citenamefont
  {Bera}, \citenamefont {Winter},\ and\ \citenamefont {Lewenstein}}]{ent3}%
  \BibitemOpen
  \bibfield  {author} {\bibinfo {author} {\bibfnamefont {A.}~\bibnamefont
  {Streltsov}}, \bibinfo {author} {\bibfnamefont {E.}~\bibnamefont
  {Chitambar}}, \bibinfo {author} {\bibfnamefont {S.}~\bibnamefont {Rana}},
  \bibinfo {author} {\bibfnamefont {M.~N.}\ \bibnamefont {Bera}}, \bibinfo
  {author} {\bibfnamefont {A.}~\bibnamefont {Winter}}, \ and\ \bibinfo {author}
  {\bibfnamefont {M.}~\bibnamefont {Lewenstein}},\ }\href {\doibase
  10.1103/PhysRevLett.116.240405} {\bibfield  {journal} {\bibinfo  {journal}
  {Phys. Rev. Lett.}\ }\textbf {\bibinfo {volume} {116}},\ \bibinfo {pages}
  {240405} (\bibinfo {year} {2016})}\BibitemShut {NoStop}%
\bibitem [{\citenamefont {{Qi}}\ \emph {et~al.}(2016)\citenamefont {{Qi}},
  \citenamefont {{Gao}},\ and\ \citenamefont {{Yan}}}]{ent4}%
  \BibitemOpen
  \bibfield  {author} {\bibinfo {author} {\bibfnamefont {X.}~\bibnamefont
  {{Qi}}}, \bibinfo {author} {\bibfnamefont {T.}~\bibnamefont {{Gao}}}, \ and\
  \bibinfo {author} {\bibfnamefont {F.}~\bibnamefont {{Yan}}},\ }\href@noop {}
  {\bibfield  {journal} {\bibinfo  {journal} {ArXiv e-prints}\ } (\bibinfo
  {year} {2016})},\ \Eprint {http://arxiv.org/abs/1610.07052} {arXiv:1610.07052
  [quant-ph]} \BibitemShut {NoStop}%
\bibitem [{\citenamefont {{Chin}}(2017)}]{ent5}%
  \BibitemOpen
  \bibfield  {author} {\bibinfo {author} {\bibfnamefont {S.}~\bibnamefont
  {{Chin}}},\ }\href@noop {} {\bibfield  {journal} {\bibinfo  {journal} {ArXiv
  e-prints}\ } (\bibinfo {year} {2017})},\ \Eprint
  {http://arxiv.org/abs/1702.03219} {arXiv:1702.03219 [quant-ph]} \BibitemShut
  {NoStop}%
\bibitem [{\citenamefont {{Zhu}}\ \emph
  {et~al.}(2017{\natexlab{a}})\citenamefont {{Zhu}}, \citenamefont {{Ma}},
  \citenamefont {{Cao}}, \citenamefont {{Fei}},\ and\ \citenamefont
  {{Vedral}}}]{ent6}%
  \BibitemOpen
  \bibfield  {author} {\bibinfo {author} {\bibfnamefont {H.}~\bibnamefont
  {{Zhu}}}, \bibinfo {author} {\bibfnamefont {Z.}~\bibnamefont {{Ma}}},
  \bibinfo {author} {\bibfnamefont {Z.}~\bibnamefont {{Cao}}}, \bibinfo
  {author} {\bibfnamefont {S.-M.}\ \bibnamefont {{Fei}}}, \ and\ \bibinfo
  {author} {\bibfnamefont {V.}~\bibnamefont {{Vedral}}},\ }\href@noop {}
  {\bibfield  {journal} {\bibinfo  {journal} {ArXiv e-prints}\ } (\bibinfo
  {year} {2017}{\natexlab{a}})},\ \Eprint {http://arxiv.org/abs/1704.01935}
  {arXiv:1704.01935 [quant-ph]} \BibitemShut {NoStop}%
\bibitem [{\citenamefont {{Zhu}}\ \emph
  {et~al.}(2017{\natexlab{b}})\citenamefont {{Zhu}}, \citenamefont
  {{Hayashi}},\ and\ \citenamefont {{Chen}}}]{ent7}%
  \BibitemOpen
  \bibfield  {author} {\bibinfo {author} {\bibfnamefont {H.}~\bibnamefont
  {{Zhu}}}, \bibinfo {author} {\bibfnamefont {M.}~\bibnamefont {{Hayashi}}}, \
  and\ \bibinfo {author} {\bibfnamefont {L.}~\bibnamefont {{Chen}}},\
  }\href@noop {} {\bibfield  {journal} {\bibinfo  {journal} {ArXiv e-prints}\ }
  (\bibinfo {year} {2017}{\natexlab{b}})},\ \Eprint
  {http://arxiv.org/abs/1704.02896} {arXiv:1704.02896 [quant-ph]} \BibitemShut
  {NoStop}%
\bibitem [{\citenamefont {Killoran}\ \emph {et~al.}(2016)\citenamefont
  {Killoran}, \citenamefont {Steinhoff},\ and\ \citenamefont
  {Plenio}}]{killoran}%
  \BibitemOpen
  \bibfield  {author} {\bibinfo {author} {\bibfnamefont {N.}~\bibnamefont
  {Killoran}}, \bibinfo {author} {\bibfnamefont {F.~E.~S.}\ \bibnamefont
  {Steinhoff}}, \ and\ \bibinfo {author} {\bibfnamefont {M.~B.}\ \bibnamefont
  {Plenio}},\ }\href {\doibase 10.1103/PhysRevLett.116.080402} {\bibfield
  {journal} {\bibinfo  {journal} {Phys. Rev. Lett.}\ }\textbf {\bibinfo
  {volume} {116}},\ \bibinfo {pages} {080402} (\bibinfo {year}
  {2016})}\BibitemShut {NoStop}%
\bibitem [{\citenamefont {Yao}\ \emph {et~al.}(2015)\citenamefont {Yao},
  \citenamefont {Xiao}, \citenamefont {Ge},\ and\ \citenamefont
  {Sun}}]{chinese}%
  \BibitemOpen
  \bibfield  {author} {\bibinfo {author} {\bibfnamefont {Y.}~\bibnamefont
  {Yao}}, \bibinfo {author} {\bibfnamefont {X.}~\bibnamefont {Xiao}}, \bibinfo
  {author} {\bibfnamefont {L.}~\bibnamefont {Ge}}, \ and\ \bibinfo {author}
  {\bibfnamefont {C.~P.}\ \bibnamefont {Sun}},\ }\href {\doibase
  10.1103/PhysRevA.92.022112} {\bibfield  {journal} {\bibinfo  {journal} {Phys.
  Rev. A}\ }\textbf {\bibinfo {volume} {92}},\ \bibinfo {pages} {022112}
  (\bibinfo {year} {2015})}\BibitemShut {NoStop}%
\bibitem [{\citenamefont {Ma}\ \emph {et~al.}(2016)\citenamefont {Ma},
  \citenamefont {Yadin}, \citenamefont {Girolami}, \citenamefont {Vedral},\
  and\ \citenamefont {Gu}}]{disc1}%
  \BibitemOpen
  \bibfield  {author} {\bibinfo {author} {\bibfnamefont {J.}~\bibnamefont
  {Ma}}, \bibinfo {author} {\bibfnamefont {B.}~\bibnamefont {Yadin}}, \bibinfo
  {author} {\bibfnamefont {D.}~\bibnamefont {Girolami}}, \bibinfo {author}
  {\bibfnamefont {V.}~\bibnamefont {Vedral}}, \ and\ \bibinfo {author}
  {\bibfnamefont {M.}~\bibnamefont {Gu}},\ }\href {\doibase
  10.1103/PhysRevLett.116.160407} {\bibfield  {journal} {\bibinfo  {journal}
  {Phys. Rev. Lett.}\ }\textbf {\bibinfo {volume} {116}},\ \bibinfo {pages}
  {160407} (\bibinfo {year} {2016})}\BibitemShut {NoStop}%
\bibitem [{\citenamefont {{Guo}}\ and\ \citenamefont
  {{Goswami}}(2016)}]{disc2}%
  \BibitemOpen
  \bibfield  {author} {\bibinfo {author} {\bibfnamefont {Y.}~\bibnamefont
  {{Guo}}}\ and\ \bibinfo {author} {\bibfnamefont {S.}~\bibnamefont
  {{Goswami}}},\ }\href@noop {} {\bibfield  {journal} {\bibinfo  {journal}
  {ArXiv e-prints}\ } (\bibinfo {year} {2016})},\ \Eprint
  {http://arxiv.org/abs/1611.00413} {arXiv:1611.00413 [quant-ph]} \BibitemShut
  {NoStop}%
\bibitem [{\citenamefont {{Mitchison}}\ \emph {et~al.}(2015)\citenamefont
  {{Mitchison}}, \citenamefont {{Woods}}, \citenamefont {{Prior}},\ and\
  \citenamefont {{Huber}}}]{huber}%
  \BibitemOpen
  \bibfield  {author} {\bibinfo {author} {\bibfnamefont {M.~T.}\ \bibnamefont
  {{Mitchison}}}, \bibinfo {author} {\bibfnamefont {M.~P.}\ \bibnamefont
  {{Woods}}}, \bibinfo {author} {\bibfnamefont {J.}~\bibnamefont {{Prior}}}, \
  and\ \bibinfo {author} {\bibfnamefont {M.}~\bibnamefont {{Huber}}},\ }\href
  {\doibase 10.1088/1367-2630/17/11/115013} {\bibfield  {journal} {\bibinfo
  {journal} {New Journal of Physics}\ }\textbf {\bibinfo {volume} {17}},\
  \bibinfo {eid} {115013} (\bibinfo {year} {2015})}\BibitemShut {NoStop}%
\bibitem [{\citenamefont {{Kammerlander}}\ and\ \citenamefont
  {{Anders}}(2016)}]{anders}%
  \BibitemOpen
  \bibfield  {author} {\bibinfo {author} {\bibfnamefont {P.}~\bibnamefont
  {{Kammerlander}}}\ and\ \bibinfo {author} {\bibfnamefont {J.}~\bibnamefont
  {{Anders}}},\ }\href {\doibase 10.1038/srep22174} {\bibfield  {journal}
  {\bibinfo  {journal} {Scientific Reports}\ }\textbf {\bibinfo {volume} {6}},\
  \bibinfo {eid} {22174} (\bibinfo {year} {2016})}\BibitemShut {NoStop}%
\bibitem [{\citenamefont {{Korzekwa}}\ \emph {et~al.}(2016)\citenamefont
  {{Korzekwa}}, \citenamefont {{Lostaglio}}, \citenamefont {{Oppenheim}},\ and\
  \citenamefont {{Jennings}}}]{lostaglio}%
  \BibitemOpen
  \bibfield  {author} {\bibinfo {author} {\bibfnamefont {K.}~\bibnamefont
  {{Korzekwa}}}, \bibinfo {author} {\bibfnamefont {M.}~\bibnamefont
  {{Lostaglio}}}, \bibinfo {author} {\bibfnamefont {J.}~\bibnamefont
  {{Oppenheim}}}, \ and\ \bibinfo {author} {\bibfnamefont {D.}~\bibnamefont
  {{Jennings}}},\ }\href {\doibase 10.1088/1367-2630/18/2/023045} {\bibfield
  {journal} {\bibinfo  {journal} {New Journal of Physics}\ }\textbf {\bibinfo
  {volume} {18}},\ \bibinfo {eid} {023045} (\bibinfo {year}
  {2016})}\BibitemShut {NoStop}%
\bibitem [{\citenamefont {Hillery}(2016)}]{hillery}%
  \BibitemOpen
  \bibfield  {author} {\bibinfo {author} {\bibfnamefont {M.}~\bibnamefont
  {Hillery}},\ }\href {\doibase 10.1103/PhysRevA.93.012111} {\bibfield
  {journal} {\bibinfo  {journal} {Phys. Rev. A}\ }\textbf {\bibinfo {volume}
  {93}},\ \bibinfo {pages} {012111} (\bibinfo {year} {2016})}\BibitemShut
  {NoStop}%
\bibitem [{\citenamefont {{Anand}}\ and\ \citenamefont {{Pati}}(2016)}]{namit}%
  \BibitemOpen
  \bibfield  {author} {\bibinfo {author} {\bibfnamefont {N.}~\bibnamefont
  {{Anand}}}\ and\ \bibinfo {author} {\bibfnamefont {A.~K.}\ \bibnamefont
  {{Pati}}},\ }\href@noop {} {\bibfield  {journal} {\bibinfo  {journal} {ArXiv
  e-prints}\ } (\bibinfo {year} {2016})},\ \Eprint
  {http://arxiv.org/abs/1611.04542} {arXiv:1611.04542 [quant-ph]} \BibitemShut
  {NoStop}%
\bibitem [{\citenamefont {Shi}\ \emph {et~al.}(2017)\citenamefont {Shi},
  \citenamefont {Liu}, \citenamefont {Wang}, \citenamefont {Yang},
  \citenamefont {Yang},\ and\ \citenamefont {Fan}}]{hengfangrover}%
  \BibitemOpen
  \bibfield  {author} {\bibinfo {author} {\bibfnamefont {H.-L.}\ \bibnamefont
  {Shi}}, \bibinfo {author} {\bibfnamefont {S.-Y.}\ \bibnamefont {Liu}},
  \bibinfo {author} {\bibfnamefont {X.-H.}\ \bibnamefont {Wang}}, \bibinfo
  {author} {\bibfnamefont {W.-L.}\ \bibnamefont {Yang}}, \bibinfo {author}
  {\bibfnamefont {Z.-Y.}\ \bibnamefont {Yang}}, \ and\ \bibinfo {author}
  {\bibfnamefont {H.}~\bibnamefont {Fan}},\ }\href {\doibase
  10.1103/PhysRevA.95.032307} {\bibfield  {journal} {\bibinfo  {journal} {Phys.
  Rev. A}\ }\textbf {\bibinfo {volume} {95}},\ \bibinfo {pages} {032307}
  (\bibinfo {year} {2017})}\BibitemShut {NoStop}%
\bibitem [{\citenamefont {Karpat}\ \emph {et~al.}(2014)\citenamefont {Karpat},
  \citenamefont {\ifmmode~\mbox{\c{C}}\else \c{C}\fi{}akmak},\ and\
  \citenamefont {Fanchini}}]{cakmak}%
  \BibitemOpen
  \bibfield  {author} {\bibinfo {author} {\bibfnamefont {G.}~\bibnamefont
  {Karpat}}, \bibinfo {author} {\bibfnamefont {B.}~\bibnamefont
  {\ifmmode~\mbox{\c{C}}\else \c{C}\fi{}akmak}}, \ and\ \bibinfo {author}
  {\bibfnamefont {F.~F.}\ \bibnamefont {Fanchini}},\ }\href {\doibase
  10.1103/PhysRevB.90.104431} {\bibfield  {journal} {\bibinfo  {journal} {Phys.
  Rev. B}\ }\textbf {\bibinfo {volume} {90}},\ \bibinfo {pages} {104431}
  (\bibinfo {year} {2014})}\BibitemShut {NoStop}%
\bibitem [{\citenamefont {Huang}\ \emph {et~al.}(2017)\citenamefont {Huang},
  \citenamefont {Situ},\ and\ \citenamefont {Zhang}}]{spin1}%
  \BibitemOpen
  \bibfield  {author} {\bibinfo {author} {\bibfnamefont {Z.}~\bibnamefont
  {Huang}}, \bibinfo {author} {\bibfnamefont {H.}~\bibnamefont {Situ}}, \ and\
  \bibinfo {author} {\bibfnamefont {C.}~\bibnamefont {Zhang}},\ }\href
  {\doibase 10.1007/s10773-017-3364-0} {\bibfield  {journal} {\bibinfo
  {journal} {International Journal of Theoretical Physics}\ ,\ \bibinfo {pages}
  {1}} (\bibinfo {year} {2017})}\BibitemShut {NoStop}%
\bibitem [{\citenamefont {Olaya-Castro}\ \emph {et~al.}(2008)\citenamefont
  {Olaya-Castro}, \citenamefont {Lee}, \citenamefont {Olsen},\ and\
  \citenamefont {Johnson}}]{bio1}%
  \BibitemOpen
  \bibfield  {author} {\bibinfo {author} {\bibfnamefont {A.}~\bibnamefont
  {Olaya-Castro}}, \bibinfo {author} {\bibfnamefont {C.~F.}\ \bibnamefont
  {Lee}}, \bibinfo {author} {\bibfnamefont {F.~F.}\ \bibnamefont {Olsen}}, \
  and\ \bibinfo {author} {\bibfnamefont {N.~F.}\ \bibnamefont {Johnson}},\
  }\href {\doibase 10.1103/PhysRevB.78.085115} {\bibfield  {journal} {\bibinfo
  {journal} {Phys. Rev. B}\ }\textbf {\bibinfo {volume} {78}},\ \bibinfo
  {pages} {085115} (\bibinfo {year} {2008})}\BibitemShut {NoStop}%
\bibitem [{\citenamefont {{Wilde}}\ \emph {et~al.}(2010)\citenamefont
  {{Wilde}}, \citenamefont {{McCracken}},\ and\ \citenamefont
  {{Mizel}}}]{bio3}%
  \BibitemOpen
  \bibfield  {author} {\bibinfo {author} {\bibfnamefont {M.~M.}\ \bibnamefont
  {{Wilde}}}, \bibinfo {author} {\bibfnamefont {J.~M.}\ \bibnamefont
  {{McCracken}}}, \ and\ \bibinfo {author} {\bibfnamefont {A.}~\bibnamefont
  {{Mizel}}},\ }\href {\doibase 10.1098/rspa.2009.0575} {\bibfield  {journal}
  {\bibinfo  {journal} {Proceedings of the Royal Society of London Series A}\
  }\textbf {\bibinfo {volume} {466}},\ \bibinfo {pages} {1347} (\bibinfo {year}
  {2010})}\BibitemShut {NoStop}%
\bibitem [{\citenamefont {Bandyopadhyay}\ \emph {et~al.}(2012)\citenamefont
  {Bandyopadhyay}, \citenamefont {Paterek},\ and\ \citenamefont
  {Kaszlikowski}}]{bio2}%
  \BibitemOpen
  \bibfield  {author} {\bibinfo {author} {\bibfnamefont {J.~N.}\ \bibnamefont
  {Bandyopadhyay}}, \bibinfo {author} {\bibfnamefont {T.}~\bibnamefont
  {Paterek}}, \ and\ \bibinfo {author} {\bibfnamefont {D.}~\bibnamefont
  {Kaszlikowski}},\ }\href {\doibase 10.1103/PhysRevLett.109.110502} {\bibfield
   {journal} {\bibinfo  {journal} {Phys. Rev. Lett.}\ }\textbf {\bibinfo
  {volume} {109}},\ \bibinfo {pages} {110502} (\bibinfo {year}
  {2012})}\BibitemShut {NoStop}%
\bibitem [{\citenamefont {{Mondal}}\ and\ \citenamefont
  {{Mukhopadhyay}}(2015)}]{chiru2}%
  \BibitemOpen
  \bibfield  {author} {\bibinfo {author} {\bibfnamefont {D.}~\bibnamefont
  {{Mondal}}}\ and\ \bibinfo {author} {\bibfnamefont {C.}~\bibnamefont
  {{Mukhopadhyay}}},\ }\href@noop {} {\bibfield  {journal} {\bibinfo  {journal}
  {ArXiv e-prints}\ } (\bibinfo {year} {2015})},\ \Eprint
  {http://arxiv.org/abs/1510.07556} {arXiv:1510.07556 [quant-ph]} \BibitemShut
  {NoStop}%
\bibitem [{\citenamefont {Wang}\ \emph {et~al.}(2016)\citenamefont {Wang},
  \citenamefont {Tian}, \citenamefont {Jing},\ and\ \citenamefont
  {Fan}}]{hengfan}%
  \BibitemOpen
  \bibfield  {author} {\bibinfo {author} {\bibfnamefont {J.}~\bibnamefont
  {Wang}}, \bibinfo {author} {\bibfnamefont {Z.}~\bibnamefont {Tian}}, \bibinfo
  {author} {\bibfnamefont {J.}~\bibnamefont {Jing}}, \ and\ \bibinfo {author}
  {\bibfnamefont {H.}~\bibnamefont {Fan}},\ }\href {\doibase
  10.1103/PhysRevA.93.062105} {\bibfield  {journal} {\bibinfo  {journal} {Phys.
  Rev. A}\ }\textbf {\bibinfo {volume} {93}},\ \bibinfo {pages} {062105}
  (\bibinfo {year} {2016})}\BibitemShut {NoStop}%
\bibitem [{\citenamefont {Giovannini}(2017)}]{curved}%
  \BibitemOpen
  \bibfield  {author} {\bibinfo {author} {\bibfnamefont {M.}~\bibnamefont
  {Giovannini}},\ }\href {http://stacks.iop.org/0264-9381/34/i=3/a=035019}
  {\bibfield  {journal} {\bibinfo  {journal} {Classical and Quantum Gravity}\
  }\textbf {\bibinfo {volume} {34}},\ \bibinfo {pages} {035019} (\bibinfo
  {year} {2017})}\BibitemShut {NoStop}%
   \bibitem [{\citenamefont {{Das}}\ and \ \citenamefont
  {{Chakrabarty}} \citenamefont
  {{Pati}} \citenamefont
  {{SenDe}} \citenamefont
  {{Sen}}(2019)}]{cohrecent1}%
  \BibitemOpen
  \bibfield  {author} {\bibinfo {author} {\bibfnamefont {S. }\  \bibnamefont
  {Das}}, \bibinfo {author}  {\bibfnamefont {I. } \bibnamefont
  {Chakraborty}}, \bibinfo {author}  {\bibfnamefont {A. K. } \bibnamefont
  {Pati}}, \bibinfo {author}  {\bibfnamefont {A.  } \bibnamefont
  {Sen(De)}},  and  \bibinfo {author}   {\bibfnamefont {U.}~\bibnamefont
  { Sen}},\ }\href {\doibase arxiv.org/abs/1812.08656} {\bibfield
  {journal} {\bibinfo{journal} {ArXiv e-prints} (\bibinfo {year} {2018}),
  }\textbf \ \bibinfo {eid} { arXiv:1812.08656 [quant-ph]  }}\BibitemShut {NoStop}%
  \bibitem [{\citenamefont {{Theurer}}\ and\ \citenamefont
  {{Egloff}} and\ \citenamefont
  {{Zhang}} and\ \citenamefont
  {{plenio}}(2019)}]{cohrecent2}%
  \BibitemOpen
  \bibfield  {author} {\bibinfo {author} {\bibfnamefont {T.}\ \bibnamefont
  {{Theurer}}}, \ \bibinfo {author} {\bibfnamefont {D.}~\bibnamefont
  {{Egloff}}},\ \bibinfo {author} {\bibfnamefont {L.}~\bibnamefont
  {{Zhang}}},\ and \bibinfo {author} {\bibfnamefont {M. B.}~\bibnamefont
  {{Plenio}}},\ }\href {\doibase 10.1103/PhysRevLett.122.190405} {\bibfield
  {journal} {\bibinfo  {journal}  {Phys. Rev. Lett.}\
  }\textbf {\bibinfo {volume} {122}},\ \bibinfo {eid} {190405} (\bibinfo {year}
  {2019})}\BibitemShut {NoStop}%
  \bibitem [{\citenamefont {{Wu}}\ \citenamefont {{Theurer}}\ and\ \citenamefont
  {{Li}} and\ \citenamefont
  {{Guo}} and \ \citenamefont
  {{Xiang}} and \ \citenamefont
  {{plenio}}(2019)}]{cohrecent3}%
  \BibitemOpen
  \bibfield  {author} {\bibinfo {author} {\bibfnamefont {K. D.}\ \bibnamefont
  {{Wu}}}, \ \bibinfo {author} {\bibfnamefont {T.}~\bibnamefont
  {{Theurer}}},\ \bibinfo {author} {\bibfnamefont {G. Y.}~\bibnamefont
  {{Xiang}}},\ \bibinfo {author} {\bibfnamefont {C. F.}~\bibnamefont
  {{Li}}},\  \bibinfo {author} {\bibfnamefont {G. C.}~\bibnamefont
  {{Guo}}}, \bibinfo {author} {\bibfnamefont {M. B.}~\bibnamefont
  {{Plenio}}}, and \bibinfo {author} {\bibfnamefont {A.}~\bibnamefont
  {{Streltsov}}},\ }\href {\doibase arxiv.org/abs/1903.01479} {\bibfield
  {journal} {\bibinfo  {journal} {ArXiv e-prints} (\bibinfo {year} {2019}),\
  }\ \bibinfo {eid} {arXiv:1903.01479 [quant-ph]}}\BibitemShut {NoStop}%
    \bibitem [{\citenamefont {{Xiong}}\ \citenamefont {{Kumar}}\ and\ \citenamefont
  {{Huang}} and\ \citenamefont
  {{Das}} and \ \citenamefont
  {{Sen}} and \ \citenamefont
  {{Wu}}(2019)}]{cohrecent4}%
  \BibitemOpen
  \bibfield  {author} {\bibinfo {author} {\bibfnamefont {C. }\ \bibnamefont
  {{Xiong}}}, \ \bibinfo {author} {\bibfnamefont {A.}~\bibnamefont
  {{Kumar}}},\ \bibinfo {author} {\bibfnamefont {M.}~\bibnamefont
  {{Huang}}},\ \bibinfo {author} {\bibfnamefont {S.}~\bibnamefont
  {{Das}}},\  \bibinfo {author} {\bibfnamefont {U.}~\bibnamefont
  {{Sen}}},\ \ and \bibinfo {author} {\bibfnamefont {J.}~\bibnamefont
  {{Wu}}},\ }\href {\doibase 10.1103/PhysRevA.99.032305} {\bibfield
  {journal} {\bibinfo  {journal} {Phys. Rev. A }\
  }\textbf {\bibinfo {volume} {99}},\ \bibinfo {eid} {032305} (\bibinfo {year} {2019})}\BibitemShut {NoStop}%
      \bibitem [{\citenamefont {{Zhengjun}}\ and \ \citenamefont
  {{Yuwen}}(2019)}]{cohrecent5}%
  \BibitemOpen
  \bibfield  {author} {\bibinfo {author} {\bibfnamefont {X.}\ \bibnamefont
  {Zhengjun}} and \ \bibinfo {author} {\bibfnamefont {S.}~\bibnamefont
  {Yuwen}},\ }\href {\doibase 10.1103/PhysRevA.99.022340} {\bibfield
  {journal} {\bibinfo  {journal} {Phys. Rev. A }\
  }\textbf {\bibinfo {volume} {99}},\ \bibinfo {eid} {022340} (\bibinfo {year} {2019})}\BibitemShut {NoStop}%
\bibitem [{\citenamefont {Scully}\ and\ \citenamefont
  {Zubairy}(1997)}]{scullybook}%
  \BibitemOpen
  \bibfield  {author} {\bibinfo {author} {\bibfnamefont {M.~O.}\ \bibnamefont
  {Scully}}\ and\ \bibinfo {author} {\bibfnamefont {M.~S.}\ \bibnamefont
  {Zubairy}},\ }\href {\doibase 10.1017/CBO9780511813993} {\emph {\bibinfo
  {title} {Quantum Optics}}}\ (\bibinfo  {publisher} {Cambridge University
  Press},\ \bibinfo {year} {1997})\BibitemShut {NoStop}%
\bibitem [{\citenamefont {{de Vicente}}\ and\ \citenamefont
  {{Streltsov}}(2017)}]{streltsov1}%
  \BibitemOpen
  \bibfield  {author} {\bibinfo {author} {\bibfnamefont {J.~I.}\ \bibnamefont
  {{de Vicente}}}\ and\ \bibinfo {author} {\bibfnamefont {A.}~\bibnamefont
  {{Streltsov}}},\ }\href {\doibase 10.1088/1751-8121/50/4/045301} {\bibfield
  {journal} {\bibinfo  {journal} {Journal of Physics A Mathematical General}\
  }\textbf {\bibinfo {volume} {50}},\ \bibinfo {eid} {045301} (\bibinfo {year}
  {2017})}\BibitemShut {NoStop}%
\bibitem [{\citenamefont {Chitambar}\ and\ \citenamefont
  {Gour}(2016{\natexlab{a}})}]{gour}%
  \BibitemOpen
  \bibfield  {author} {\bibinfo {author} {\bibfnamefont {E.}~\bibnamefont
  {Chitambar}}\ and\ \bibinfo {author} {\bibfnamefont {G.}~\bibnamefont
  {Gour}},\ }\href {\doibase 10.1103/PhysRevLett.117.030401} {\bibfield
  {journal} {\bibinfo  {journal} {Phys. Rev. Lett.}\ }\textbf {\bibinfo
  {volume} {117}},\ \bibinfo {pages} {030401} (\bibinfo {year}
  {2016}{\natexlab{a}})}\BibitemShut {NoStop}%
\bibitem [{\citenamefont {Streltsov}\ \emph
  {et~al.}(2017{\natexlab{a}})\citenamefont {Streltsov}, \citenamefont {Rana},
  \citenamefont {Bera},\ and\ \citenamefont {Lewenstein}}]{streltsov2}%
  \BibitemOpen
  \bibfield  {author} {\bibinfo {author} {\bibfnamefont {A.}~\bibnamefont
  {Streltsov}}, \bibinfo {author} {\bibfnamefont {S.}~\bibnamefont {Rana}},
  \bibinfo {author} {\bibfnamefont {M.~N.}\ \bibnamefont {Bera}}, \ and\
  \bibinfo {author} {\bibfnamefont {M.}~\bibnamefont {Lewenstein}},\ }\href
  {\doibase 10.1103/PhysRevX.7.011024} {\bibfield  {journal} {\bibinfo
  {journal} {Phys. Rev. X}\ }\textbf {\bibinfo {volume} {7}},\ \bibinfo {pages}
  {011024} (\bibinfo {year} {2017}{\natexlab{a}})}\BibitemShut {NoStop}%
  \bibitem [{\citenamefont {{Theurer}}\ \emph {et~al.}(2017)\citenamefont
  {{Theurer}}, \citenamefont {{Killoran}}, \citenamefont {{Egloff}},\ and\
  \citenamefont {{Plenio}}}]{plenio2}%
  \BibitemOpen
  \bibfield  {author} {\bibinfo {author} {\bibfnamefont {T.}~\bibnamefont
  {{Theurer}}}, \bibinfo {author} {\bibfnamefont {N.}~\bibnamefont
  {{Killoran}}}, \bibinfo {author} {\bibfnamefont {D.}~\bibnamefont
  {{Egloff}}}, \ and\ \bibinfo {author} {\bibfnamefont {M.~B.}\ \bibnamefont
  {{Plenio}}},\ }\href@noop {} \href {\doibase
10.1103/PhysRevLett.119.230401} {\bibfield  {journal} {\bibinfo  {journal} {Phys.
  Rev. Lett.}\ }\textbf {\bibinfo {volume} {119}},\ \bibinfo {pages} {230401}
  (\bibinfo {year} {2017})} \BibitemShut
    {NoStop}%
\bibitem [{\citenamefont {Singh}\ \emph {et~al.}(2015)\citenamefont {Singh},
  \citenamefont {Bera}, \citenamefont {Dhar},\ and\ \citenamefont
  {Pati}}]{mcms}%
  \BibitemOpen
  \bibfield  {author} {\bibinfo {author} {\bibfnamefont {U.}~\bibnamefont
  {Singh}}, \bibinfo {author} {\bibfnamefont {M.~N.}\ \bibnamefont {Bera}},
  \bibinfo {author} {\bibfnamefont {H.~S.}\ \bibnamefont {Dhar}}, \ and\
  \bibinfo {author} {\bibfnamefont {A.~K.}\ \bibnamefont {Pati}},\ }\href
  {\doibase 10.1103/PhysRevA.91.052115} {\bibfield  {journal} {\bibinfo
  {journal} {Phys. Rev. A}\ }\textbf {\bibinfo {volume} {91}},\ \bibinfo
  {pages} {052115} (\bibinfo {year} {2015})}\BibitemShut {NoStop}%
\bibitem [{\citenamefont {Misra}\ \emph {et~al.}(2016)\citenamefont {Misra},
  \citenamefont {Singh}, \citenamefont {Bhattacharya},\ and\ \citenamefont
  {Pati}}]{samya}%
  \BibitemOpen
  \bibfield  {author} {\bibinfo {author} {\bibfnamefont {A.}~\bibnamefont
  {Misra}}, \bibinfo {author} {\bibfnamefont {U.}~\bibnamefont {Singh}},
  \bibinfo {author} {\bibfnamefont {S.}~\bibnamefont {Bhattacharya}}, \ and\
  \bibinfo {author} {\bibfnamefont {A.~K.}\ \bibnamefont {Pati}},\ }\href
  {\doibase 10.1103/PhysRevA.93.052335} {\bibfield  {journal} {\bibinfo
  {journal} {Phys. Rev. A}\ }\textbf {\bibinfo {volume} {93}},\ \bibinfo
  {pages} {052335} (\bibinfo {year} {2016})}\BibitemShut {NoStop}%
  \bibitem [{\citenamefont {Rastegin}(2018)}]{Rastegin} \BibitemOpen \bibfield {author} {\bibinfo {author} {\bibfnamefont {A. E.}~\bibnamefont {Rastegin}},\ }\href {\doibase 10.1088/1751-8121/aab348} {\bibfield {journal} {\bibinfo {journal} { J. Phys. A: Math. Theor.}\ }\textbf {\bibinfo {volume} {51}},\ \bibinfo {pages} {414011} (\bibinfo {year} {2018}{\natexlab{b}})}\BibitemShut {NoStop}%
  \bibitem [{\citenamefont {Bischof1}\ \emph
 {et~al.}(2019{\natexlab{b}})\citenamefont {Bischof1}, \citenamefont {Kampermann},
 \ and\ \citenamefont {Bruss}}]{Bischof1}%
 \BibitemOpen \bibfield {author} {\bibinfo {author} {\bibfnamefont {F.}~\bibnamefont {Bischof}}, \bibinfo {author} {\bibfnamefont {H.}~\bibnamefont {Kampermann}}, \ and\ \bibinfo {author} {\bibfnamefont {D.}~\bibnamefont {Bru{\ss}}},\ }\href {\doibase 10.1103/PhysRevLett.123.110402} {\bibfield {journal} {\bibinfo {journal} {Phys. Rev. Lett.}\ }\textbf {\bibinfo {volume} {123}},\ \bibinfo {pages} {110402} (\bibinfo {year} {2019}{\natexlab{b}})}\BibitemShut {NoStop}%
 \bibitem [{\citenamefont {Bischof2}\ \emph
 {et~al.}(2019{\natexlab{b}})\citenamefont {Bischof2}, \citenamefont {Kampermann},
 \ and\ \citenamefont {Bruss}}]{Bischof2}%
 \BibitemOpen \bibfield {author} {\bibinfo {author} {\bibfnamefont {F.}~\bibnamefont {Bischof}}, \bibinfo {author} {\bibfnamefont {H.}~\bibnamefont {Kampermann}}, \ and\ \bibinfo {author} {\bibfnamefont {D.}~\bibnamefont {Bru{\ss}}},\ }\href {\doibase arxiv.org/abs/1907.08574} {\bibfield {journal} {\bibinfo {journal} {Arxiv e-prints}}, \bibinfo {pages} {arXiv:1907.08574 [quant-ph]}\  (\bibinfo {year} {2019}{\natexlab{b}})}\BibitemShut {NoStop}%
 \bibitem [{\citenamefont {Gisisn}(1993)}]{Gisin}%
  \BibitemOpen
  \bibfield  {author} {\bibinfo {author} {\bibfnamefont {N.}\ \bibnamefont
  {Gisin}},\ }\href {\doibase 10.1103/PhysRevLett.52.1657} {\bibfield  {journal}
  {\bibinfo  {journal} {Phys. Rev. Lett.}\ }\textbf {\bibinfo {volume} {12}},\
  \bibinfo {pages} {1657} (\bibinfo {year} {1993})}\BibitemShut {NoStop}%
 \bibitem [{\citenamefont {}\ \emph
 {et~al.}(2019{\natexlab{b}})\citenamefont {Hughston}, \citenamefont {Jozsa},
 \ and\ \citenamefont {Wooters}}]{Hughston}%
 \BibitemOpen \bibfield {author} {\bibinfo {author} {\bibfnamefont {L. P.}~\bibnamefont {Hughston}}, \bibinfo {author} {\bibfnamefont {R.}~\bibnamefont {Jozsa}}, \ and\ \bibinfo {author} {\bibfnamefont {W. K.}~\bibnamefont {Wootters}},\ }\href {\doibase 10.1016/0375-9601(93)90880-9} {\bibfield {journal} {\bibinfo {journal} {Phys. Lett. A}\ }\textbf {\bibinfo {volume} {184}},\ \bibinfo {pages} {14} (\bibinfo {year} {1993}{\natexlab{b}})}\BibitemShut {NoStop}%
\bibitem [{\citenamefont {Chitambar}\ and\ \citenamefont
  {Gour}(2016{\natexlab{b}})}]{chitambar}%
  \BibitemOpen
  \bibfield  {author} {\bibinfo {author} {\bibfnamefont {E.}~\bibnamefont
  {Chitambar}}\ and\ \bibinfo {author} {\bibfnamefont {G.}~\bibnamefont
  {Gour}},\ }\href {\doibase 10.1103/PhysRevLett.117.030401} {\bibfield
  {journal} {\bibinfo  {journal} {Phys. Rev. Lett.}\ }\textbf {\bibinfo
  {volume} {117}},\ \bibinfo {pages} {030401} (\bibinfo {year}
  {2016}{\natexlab{b}})}\BibitemShut {NoStop}%
\bibitem [{\citenamefont {Streltsov}\ \emph
  {et~al.}(2017{\natexlab{b}})\citenamefont {Streltsov}, \citenamefont {Rana},
  \citenamefont {Bera},\ and\ \citenamefont {Lewenstein}}]{streltsov}%
  \BibitemOpen
  \bibfield  {author} {\bibinfo {author} {\bibfnamefont {A.}~\bibnamefont
  {Streltsov}}, \bibinfo {author} {\bibfnamefont {S.}~\bibnamefont {Rana}},
  \bibinfo {author} {\bibfnamefont {M.~N.}\ \bibnamefont {Bera}}, \ and\
  \bibinfo {author} {\bibfnamefont {M.}~\bibnamefont {Lewenstein}},\ }\href
  {\doibase 10.1103/PhysRevX.7.011024} {\bibfield  {journal} {\bibinfo
  {journal} {Phys. Rev. X}\ }\textbf {\bibinfo {volume} {7}},\ \bibinfo {pages}
  {011024} (\bibinfo {year} {2017}{\natexlab{b}})}\BibitemShut {NoStop}%
\bibitem [{\citenamefont {Kraus}(1983)}]{kraus}%
  \BibitemOpen
  \bibfield  {author} {\bibinfo {author} {\bibfnamefont {K.}~\bibnamefont
  {Kraus}},\ }\href {https://books.google.co.in/books?id=fRBBAQAAIAAJ} {\emph
  {\bibinfo {title} {States, effects, and operations: Fundamental notions of
  quantum theory, Lectures in Mathematical Physics at the University of Texas
  at Austin}}},\ Lecture notes in Physics\ (\bibinfo  {publisher} {Springer,
  Berlin},\ \bibinfo {year} {1983})\BibitemShut {NoStop}%
\bibitem [{\citenamefont {Duan}\ and\ \citenamefont
  {Guo}(1998{\natexlab{a}})}]{duanpla}%
  \BibitemOpen
  \bibfield  {author} {\bibinfo {author} {\bibfnamefont {L.-M.}\ \bibnamefont
  {Duan}}\ and\ \bibinfo {author} {\bibfnamefont {G.-C.}\ \bibnamefont {Guo}},\
  }\href {\doibase http://dx.doi.org/10.1016/S0375-9601(98)00287-4} {\bibfield
  {journal} {\bibinfo  {journal} {Physics Letters A}\ }\textbf {\bibinfo
  {volume} {243}},\ \bibinfo {pages} {261 } (\bibinfo {year}
  {1998}{\natexlab{a}})}\BibitemShut {NoStop}%
\bibitem [{\citenamefont {Duan}\ and\ \citenamefont
  {Guo}(1998{\natexlab{b}})}]{probclon1}%
  \BibitemOpen
  \bibfield  {author} {\bibinfo {author} {\bibfnamefont {L.-M.}\ \bibnamefont
  {Duan}}\ and\ \bibinfo {author} {\bibfnamefont {G.-C.}\ \bibnamefont {Guo}},\
  }\href {\doibase 10.1103/PhysRevLett.80.4999} {\bibfield  {journal} {\bibinfo
   {journal} {Phys. Rev. Lett.}\ }\textbf {\bibinfo {volume} {80}},\ \bibinfo
  {pages} {4999} (\bibinfo {year} {1998}{\natexlab{b}})}\BibitemShut {NoStop}%
\bibitem [{\citenamefont {Pati}(1999)}]{probclon2}%
  \BibitemOpen
  \bibfield  {author} {\bibinfo {author} {\bibfnamefont {A.~K.}\ \bibnamefont
  {Pati}},\ }\href {\doibase 10.1103/PhysRevLett.83.2849} {\bibfield  {journal}
  {\bibinfo  {journal} {Phys. Rev. Lett.}\ }\textbf {\bibinfo {volume} {83}},\
  \bibinfo {pages} {2849} (\bibinfo {year} {1999})}\BibitemShut {NoStop}%
\bibitem [{\citenamefont {Nielsen}\ and\ \citenamefont
  {Chuang}(2011)}]{nielsenbook}%
  \BibitemOpen
  \bibfield  {author} {\bibinfo {author} {\bibfnamefont {M.~A.}\ \bibnamefont
  {Nielsen}}\ and\ \bibinfo {author} {\bibfnamefont {I.~L.}\ \bibnamefont
  {Chuang}},\ }\href@noop {} {\emph {\bibinfo {title} {Quantum Computation and
  Quantum Information: 10th Anniversary Edition}}},\ \bibinfo {edition} {10th}\
  ed.\ (\bibinfo  {publisher} {Cambridge University Press},\ \bibinfo {address}
  {New York, NY, USA},\ \bibinfo {year} {2011})\BibitemShut {NoStop}%
\bibitem [{\citenamefont {Rana}\ \emph {et~al.}(2016)\citenamefont {Rana},
  \citenamefont {Parashar},\ and\ \citenamefont {Lewenstein}}]{swapan}%
  \BibitemOpen
  \bibfield  {author} {\bibinfo {author} {\bibfnamefont {S.}~\bibnamefont
  {Rana}}, \bibinfo {author} {\bibfnamefont {P.}~\bibnamefont {Parashar}}, \
  and\ \bibinfo {author} {\bibfnamefont {M.}~\bibnamefont {Lewenstein}},\
  }\href {\doibase 10.1103/PhysRevA.93.012110} {\bibfield  {journal} {\bibinfo
  {journal} {Phys. Rev. A}\ }\textbf {\bibinfo {volume} {93}},\ \bibinfo
  {pages} {012110} (\bibinfo {year} {2016})}\BibitemShut {NoStop}%
\bibitem [{\citenamefont {Rastegin}(2016)}]{rastegin}%
  \BibitemOpen
  \bibfield  {author} {\bibinfo {author} {\bibfnamefont {A.~E.}\ \bibnamefont
  {Rastegin}},\ }\href {\doibase 10.1103/PhysRevA.93.032136} {\bibfield
  {journal} {\bibinfo  {journal} {Phys. Rev. A}\ }\textbf {\bibinfo {volume}
  {93}},\ \bibinfo {pages} {032136} (\bibinfo {year} {2016})}\BibitemShut
  {NoStop}%
\bibitem [{\citenamefont {{Shao}}\ \emph {et~al.}(2016)\citenamefont {{Shao}},
  \citenamefont {{Li}}, \citenamefont {{Luo}},\ and\ \citenamefont
  {{Xi}}}]{renyicoh}%
  \BibitemOpen
  \bibfield  {author} {\bibinfo {author} {\bibfnamefont {L.-H.}\ \bibnamefont
  {{Shao}}}, \bibinfo {author} {\bibfnamefont {Y.}~\bibnamefont {{Li}}},
  \bibinfo {author} {\bibfnamefont {Y.}~\bibnamefont {{Luo}}}, \ and\ \bibinfo
  {author} {\bibfnamefont {Z.}~\bibnamefont {{Xi}}},\ }\href@noop {} {\bibfield
   {journal} {\bibinfo  {journal} {ArXiv e-prints}\ } (\bibinfo {year}
  {2016})},\ \Eprint {http://arxiv.org/abs/1609.08759} {arXiv:1609.08759
  [quant-ph]} \BibitemShut {NoStop}%
\bibitem [{\citenamefont {Horodecki}\ \emph {et~al.}(2009)\citenamefont
  {Horodecki}, \citenamefont {Horodecki}, \citenamefont {Horodecki},\ and\
  \citenamefont {Horodecki}}]{horodeckireview}%
  \BibitemOpen
  \bibfield  {author} {\bibinfo {author} {\bibfnamefont {R.}~\bibnamefont
  {Horodecki}}, \bibinfo {author} {\bibfnamefont {P.}~\bibnamefont
  {Horodecki}}, \bibinfo {author} {\bibfnamefont {M.}~\bibnamefont
  {Horodecki}}, \ and\ \bibinfo {author} {\bibfnamefont {K.}~\bibnamefont
  {Horodecki}},\ }\href {\doibase 10.1103/RevModPhys.81.865} {\bibfield
  {journal} {\bibinfo  {journal} {Rev. Mod. Phys.}\ }\textbf {\bibinfo {volume}
  {81}},\ \bibinfo {pages} {865} (\bibinfo {year} {2009})}\BibitemShut
  {NoStop}%
\bibitem [{\citenamefont {{Streltsov}}\ \emph {et~al.}(2016)\citenamefont
  {{Streltsov}}, \citenamefont {{Adesso}},\ and\ \citenamefont
  {{Plenio}}}]{adessorev}%
  \BibitemOpen
  \bibfield  {author} {\bibinfo {author} {\bibfnamefont {A.}~\bibnamefont
  {{Streltsov}}}, \bibinfo {author} {\bibfnamefont {G.}~\bibnamefont
  {{Adesso}}}, \ and\ \bibinfo {author} {\bibfnamefont {M.~B.}\ \bibnamefont
  {{Plenio}}},\ }\href@noop {} {\bibfield  {journal} {\bibinfo  {journal}
  {ArXiv e-prints}\ } (\bibinfo {year} {2016})},\ \Eprint
  {http://arxiv.org/abs/1609.02439} {arXiv:1609.02439 [quant-ph]} \BibitemShut
  {NoStop}%
\bibitem [{\citenamefont {{Horodecki}}\ and\ \citenamefont
  {{Oppenheim}}(2013)}]{nanoscale}%
  \BibitemOpen
  \bibfield  {author} {\bibinfo {author} {\bibfnamefont {M.}~\bibnamefont
  {{Horodecki}}}\ and\ \bibinfo {author} {\bibfnamefont {J.}~\bibnamefont
  {{Oppenheim}}},\ }\href {\doibase 10.1038/ncomms3059} {\bibfield  {journal}
  {\bibinfo  {journal} {Nature Communications}\ }\textbf {\bibinfo {volume}
  {4}},\ \bibinfo {eid} {2059} (\bibinfo {year} {2013})}\BibitemShut {NoStop}%
\bibitem [{\citenamefont {Brand\~ao}\ \emph {et~al.}(2013)\citenamefont
  {Brand\~ao}, \citenamefont {Horodecki}, \citenamefont {Oppenheim},
  \citenamefont {Renes},\ and\ \citenamefont {Spekkens}}]{brandao}%
  \BibitemOpen
  \bibfield  {author} {\bibinfo {author} {\bibfnamefont {F.~G. S.~L.}\
  \bibnamefont {Brand\~ao}}, \bibinfo {author} {\bibfnamefont {M.}~\bibnamefont
  {Horodecki}}, \bibinfo {author} {\bibfnamefont {J.}~\bibnamefont
  {Oppenheim}}, \bibinfo {author} {\bibfnamefont {J.~M.}\ \bibnamefont
  {Renes}}, \ and\ \bibinfo {author} {\bibfnamefont {R.~W.}\ \bibnamefont
  {Spekkens}},\ }\href {\doibase 10.1103/PhysRevLett.111.250404} {\bibfield
  {journal} {\bibinfo  {journal} {Phys. Rev. Lett.}\ }\textbf {\bibinfo
  {volume} {111}},\ \bibinfo {pages} {250404} (\bibinfo {year}
  {2013})}\BibitemShut {NoStop}%
\bibitem [{\citenamefont {Feynman}\ \emph {et~al.}(1963)\citenamefont
  {Feynman}, \citenamefont {Leighton},\ and\ \citenamefont
  {Sands}}]{double_slit_book}%
  \BibitemOpen
  \bibfield  {author} {\bibinfo {author} {\bibfnamefont {R.}~\bibnamefont
  {Feynman}}, \bibinfo {author} {\bibfnamefont {R.}~\bibnamefont {Leighton}}, \
  and\ \bibinfo {author} {\bibfnamefont {M.}~\bibnamefont {Sands}},\ }\href
  {https://books.google.co.in/books?id=\_6XvAAAAMAAJ} {\emph {\bibinfo {title}
  {The Feynman Lectures on Physics}}},\ \bibinfo {series} {The Feynman Lectures
  on Physics}\ No.\ \bibinfo {number} {v. 3}\ (\bibinfo  {publisher}
  {Pearson/Addison-Wesley},\ \bibinfo {year} {1963})\BibitemShut {NoStop}%
\bibitem [{\citenamefont {Englert}(1996)}]{englert1}%
  \BibitemOpen
  \bibfield  {author} {\bibinfo {author} {\bibfnamefont {B.-G.}\ \bibnamefont
  {Englert}},\ }\href {\doibase 10.1103/PhysRevLett.77.2154} {\bibfield
  {journal} {\bibinfo  {journal} {Phys. Rev. Lett.}\ }\textbf {\bibinfo
  {volume} {77}},\ \bibinfo {pages} {2154} (\bibinfo {year}
  {1996})}\BibitemShut {NoStop}%
\bibitem [{\citenamefont {{Greenberger}}\ and\ \citenamefont
  {{Yasin}}(1988)}]{GREENBERGER1988391}%
  \BibitemOpen
  \bibfield  {author} {\bibinfo {author} {\bibfnamefont {D.~M.}\ \bibnamefont
  {{Greenberger}}}\ and\ \bibinfo {author} {\bibfnamefont {A.}~\bibnamefont
  {{Yasin}}},\ }\href {\doibase 10.1016/0375-9601(88)90114-4} {\bibfield
  {journal} {\bibinfo  {journal} {Physics Letters A}\ }\textbf {\bibinfo
  {volume} {128}},\ \bibinfo {pages} {391} (\bibinfo {year}
  {1988})}\BibitemShut {NoStop}%
\bibitem [{\citenamefont {Jaeger}\ \emph {et~al.}(1995)\citenamefont {Jaeger},
  \citenamefont {Shimony},\ and\ \citenamefont {Vaidman}}]{Jaeger-shimony}%
  \BibitemOpen
  \bibfield  {author} {\bibinfo {author} {\bibfnamefont {G.}~\bibnamefont
  {Jaeger}}, \bibinfo {author} {\bibfnamefont {A.}~\bibnamefont {Shimony}}, \
  and\ \bibinfo {author} {\bibfnamefont {L.}~\bibnamefont {Vaidman}},\ }\href
  {\doibase 10.1103/PhysRevA.51.54} {\bibfield  {journal} {\bibinfo  {journal}
  {Phys. Rev. A}\ }\textbf {\bibinfo {volume} {51}},\ \bibinfo {pages} {54}
  (\bibinfo {year} {1995})}\BibitemShut {NoStop}%
\bibitem [{\citenamefont {Wootters}\ and\ \citenamefont {Zurek}(1979)}]{Zurek}%
  \BibitemOpen
  \bibfield  {author} {\bibinfo {author} {\bibfnamefont {W.~K.}\ \bibnamefont
  {Wootters}}\ and\ \bibinfo {author} {\bibfnamefont {W.~H.}\ \bibnamefont
  {Zurek}},\ }\href {\doibase 10.1103/PhysRevD.19.473} {\bibfield  {journal}
  {\bibinfo  {journal} {Phys. Rev. D}\ }\textbf {\bibinfo {volume} {19}},\
  \bibinfo {pages} {473} (\bibinfo {year} {1979})}\BibitemShut {NoStop}%
\bibitem [{\citenamefont {Englert}\ \emph {et~al.}(1992)\citenamefont
  {Englert}, \citenamefont {Walther},\ and\ \citenamefont
  {Scully}}]{Englert1992}%
  \BibitemOpen
  \bibfield  {author} {\bibinfo {author} {\bibfnamefont {B.-G.}\ \bibnamefont
  {Englert}}, \bibinfo {author} {\bibfnamefont {H.}~\bibnamefont {Walther}}, \
  and\ \bibinfo {author} {\bibfnamefont {M.~O.}\ \bibnamefont {Scully}},\
  }\href {\doibase 10.1007/BF00325381} {\bibfield  {journal} {\bibinfo
  {journal} {Applied Physics B}\ }\textbf {\bibinfo {volume} {54}},\ \bibinfo
  {pages} {366} (\bibinfo {year} {1992})}\BibitemShut {NoStop}%
\bibitem [{\citenamefont {Mandel}(1991)}]{Mandel:91}%
  \BibitemOpen
  \bibfield  {author} {\bibinfo {author} {\bibfnamefont {L.}~\bibnamefont
  {Mandel}},\ }\href {\doibase 10.1364/OL.16.001882} {\bibfield  {journal}
  {\bibinfo  {journal} {Opt. Lett.}\ }\textbf {\bibinfo {volume} {16}},\
  \bibinfo {pages} {1882} (\bibinfo {year} {1991})}\BibitemShut {NoStop}%
\bibitem [{\citenamefont {Bera}\ \emph {et~al.}(2015)\citenamefont {Bera},
  \citenamefont {Qureshi}, \citenamefont {Siddiqui},\ and\ \citenamefont
  {Pati}}]{Manab}%
  \BibitemOpen
  \bibfield  {author} {\bibinfo {author} {\bibfnamefont {M.~N.}\ \bibnamefont
  {Bera}}, \bibinfo {author} {\bibfnamefont {T.}~\bibnamefont {Qureshi}},
  \bibinfo {author} {\bibfnamefont {M.~A.}\ \bibnamefont {Siddiqui}}, \ and\
  \bibinfo {author} {\bibfnamefont {A.~K.}\ \bibnamefont {Pati}},\ }\href
  {\doibase 10.1103/PhysRevA.92.012118} {\bibfield  {journal} {\bibinfo
  {journal} {Phys. Rev. A}\ }\textbf {\bibinfo {volume} {92}},\ \bibinfo
  {pages} {012118} (\bibinfo {year} {2015})}\BibitemShut {NoStop}%
\bibitem [{\citenamefont {Duan}\ and\ \citenamefont
  {Guo}(1998{\natexlab{c}})}]{DuanGuoPRL}%
  \BibitemOpen
  \bibfield  {author} {\bibinfo {author} {\bibfnamefont {L.-M.}\ \bibnamefont
  {Duan}}\ and\ \bibinfo {author} {\bibfnamefont {G.-C.}\ \bibnamefont {Guo}},\
  }\href {\doibase 10.1103/PhysRevLett.80.4999} {\bibfield  {journal} {\bibinfo
   {journal} {Phys. Rev. Lett.}\ }\textbf {\bibinfo {volume} {80}},\ \bibinfo
  {pages} {4999} (\bibinfo {year} {1998}{\natexlab{c}})}\BibitemShut {NoStop}%
\bibitem [{\citenamefont {Ivanovic}(1987)}]{IVANOVIC1987257}%
  \BibitemOpen
  \bibfield  {author} {\bibinfo {author} {\bibfnamefont {I.}~\bibnamefont
  {Ivanovic}},\ }\href {\doibase https://doi.org/10.1016/0375-9601(87)90222-2}
  {\bibfield  {journal} {\bibinfo  {journal} {Phys. Lett. A}\ }\textbf
  {\bibinfo {volume} {123}},\ \bibinfo {pages} {257 } (\bibinfo {year}
  {1987})}\BibitemShut {NoStop}%
\bibitem [{\citenamefont {{Chefles}}\ and\ \citenamefont
  {{Barnett}}(1998)}]{CHEFLES1998339}%
  \BibitemOpen
  \bibfield  {author} {\bibinfo {author} {\bibfnamefont {A.}~\bibnamefont
  {{Chefles}}}\ and\ \bibinfo {author} {\bibfnamefont {S.~M.}\ \bibnamefont
  {{Barnett}}},\ }\href {\doibase 10.1088/0305-4470/31/50/007} {\bibfield
  {journal} {\bibinfo  {journal} {J. Phys. A Math. Gen.}\ }\textbf {\bibinfo
  {volume} {31}},\ \bibinfo {pages} {10097} (\bibinfo {year} {1998})},\ \Eprint
  {http://arxiv.org/abs/quant-ph/9808018} {quant-ph/9808018} \BibitemShut
  {NoStop}%
\bibitem [{\citenamefont {{Chefles}}(1998)}]{0305-4470-31-50-007}%
  \BibitemOpen
  \bibfield  {author} {\bibinfo {author} {\bibfnamefont {A.}~\bibnamefont
  {{Chefles}}},\ }\href {\doibase 10.1016/S0375-9601(98)00064-4} {\bibfield
  {journal} {\bibinfo  {journal} {Phys. Lett. A}\ }\textbf {\bibinfo {volume}
  {239}},\ \bibinfo {pages} {339} (\bibinfo {year} {1998})},\ \Eprint
  {http://arxiv.org/abs/quant-ph/9807022} {quant-ph/9807022} \BibitemShut
  {NoStop}%
\bibitem [{\citenamefont {Barnett}\ and\ \citenamefont
  {Croke}(2009)}]{Barnett:09}%
  \BibitemOpen
  \bibfield  {author} {\bibinfo {author} {\bibfnamefont {S.~M.}\ \bibnamefont
  {Barnett}}\ and\ \bibinfo {author} {\bibfnamefont {S.}~\bibnamefont
  {Croke}},\ }\href {\doibase 10.1364/AOP.1.000238} {\bibfield  {journal}
  {\bibinfo  {journal} {Adv. Opt. Photon.}\ }\textbf {\bibinfo {volume} {1}},\
  \bibinfo {pages} {238} (\bibinfo {year} {2009})}\BibitemShut {NoStop}%
\bibitem [{\citenamefont {Dieks}(1988)}]{Dieks}%
  \BibitemOpen
  \bibfield  {author} {\bibinfo {author} {\bibfnamefont {D.}~\bibnamefont
  {Dieks}},\ }\href {\doibase https://doi.org/10.1016/0375-9601(88)90840-7}
  {\bibfield  {journal} {\bibinfo  {journal} {Phys. Lett. A}\ }\textbf
  {\bibinfo {volume} {126}},\ \bibinfo {pages} {303 } (\bibinfo {year}
  {1988})}\BibitemShut {NoStop}%
\bibitem [{\citenamefont {Jaeger}\ and\ \citenamefont
  {Shimony}(1995)}]{Jaeger}%
  \BibitemOpen
  \bibfield  {author} {\bibinfo {author} {\bibfnamefont {G.}~\bibnamefont
  {Jaeger}}\ and\ \bibinfo {author} {\bibfnamefont {A.}~\bibnamefont
  {Shimony}},\ }\href {\doibase https://doi.org/10.1016/0375-9601(94)00919-G}
  {\bibfield  {journal} {\bibinfo  {journal} {Phys. Lett. A}\ }\textbf
  {\bibinfo {volume} {197}},\ \bibinfo {pages} {83 } (\bibinfo {year}
  {1995})}\BibitemShut {NoStop}%
\bibitem [{\citenamefont {Peres}(1988)}]{Peres}%
  \BibitemOpen
  \bibfield  {author} {\bibinfo {author} {\bibfnamefont {A.}~\bibnamefont
  {Peres}},\ }\href {\doibase https://doi.org/10.1016/0375-9601(88)91034-1}
  {\bibfield  {journal} {\bibinfo  {journal} {Physics Letters A}\ }\textbf
  {\bibinfo {volume} {128}},\ \bibinfo {pages} {19} (\bibinfo {year}
  {1988})}\BibitemShut {NoStop}%
  \bibitem [{\citenamefont {Sakurai}\ and\ \citenamefont {Napolitano}(2013)}]{sakuraibook} \BibitemOpen \bibfield {author} {\bibinfo {author} {\bibfnamefont {J.~J.}\ \bibnamefont {Sakurai}}\ and\ \bibinfo {author} {\bibfnamefont {J.}\ \bibnamefont {Napolitano}},\ }\href@noop {} {\emph {\bibinfo {title} {Modern Quantum Mechanics}}},\ \bibinfo {edition} {2nd}\ ed.\ (\bibinfo {publisher} {Cambridge University Press},\ \bibinfo {address} {New York, NY, USA},\ \bibinfo {year} {2013})\BibitemShut {NoStop}%
  \bibitem [{\citenamefont {pati}\ and\ \citenamefont {Maccone}(2014)}]{Pati} \BibitemOpen \bibfield {author} {\bibinfo {author} {\bibfnamefont {L.}~\bibnamefont {Maccone}\ and\ \bibinfo {author} {\bibfnamefont {A.}~\bibnamefont {Pati}},\ }\href {\doibase https://doi.org/10.1103/PhysRevLett.113.260401} {\bibfield {journal} {\bibinfo {journal} Phys. Rev. Lett. }\ }\textbf {\bibinfo {volume} {113}},\ \bibinfo {pages} {260401} (\bibinfo {year} {2014})}\BibitemShut {NoStop}%
  \bibitem [{\citenamefont {Yao}\ \emph {et~al.}(2015)\citenamefont {Yao}, \citenamefont {Xiao}, \citenamefont {Wang},\ and\ \citenamefont {Pati}}]{Sun} \BibitemOpen \bibfield {author} {\bibinfo {author} {\bibfnamefont {Y.}\ \bibnamefont {Yao}}, \bibinfo {author} {\bibfnamefont {X.}~\bibnamefont {Xiao}}, \bibinfo {author} {\bibfnamefont {X.}\ \bibnamefont {Wang}}, \ and\ \bibinfo {author} {\bibfnamefont {C. P.}\ \bibnamefont {Sun}},\ }\href {\doibase 10.1103/PhysRevA.91.062113} {\bibfield {journal} {\bibinfo {journal} {Phys. Rev. A}\ }\textbf {\bibinfo {volume} {91}},\ \bibinfo {pages} {062113} (\bibinfo {year} {2015})}\BibitemShut {NoStop}%
  \bibitem [{\citenamefont {Xiao}\ \emph {et~al.}(2016) \citenamefont {Xiao}, \citenamefont { Jing,},\citenamefont {Jost},\ and\ \citenamefont {Fei}}]{Fei} \BibitemOpen \bibfield {author} {\bibinfo {author} {\bibfnamefont {Y.}\ \bibnamefont ~\bibnamefont {Xiao}}, \bibinfo {author} {\bibfnamefont {N.}\ \bibnamefont {Jing}}, \bibinfo {author} {\bibfnamefont {X. L.}\ \bibnamefont {Jost}}, \ and\ \bibinfo {author} {\bibfnamefont {S. M.}\ \bibnamefont {Fei}},\ }\href {\doibase 10.1038/srep23201} {\bibfield {journal} {\bibinfo {journal} {Sci Rep}\ }\textbf {\bibinfo {volume} {6}},\ \bibinfo {pages} {23201} (\bibinfo {year} {2016})}\BibitemShut {NoStop}%
\bibitem [{\citenamefont {Wang}\ \emph {et~al.}(2016) \citenamefont {Wang}, \citenamefont { Zhan},\citenamefont {Bian},\citenamefont {Li},\citenamefont {Zhang}, \ and\ \citenamefont {Xue}}] {Xue} \BibitemOpen \bibfield {author} {\bibinfo {author} {\bibfnamefont {K.}\ \bibnamefont ~\bibnamefont {Wang}}, \bibinfo {author} {\bibfnamefont {X.}\ \bibnamefont {Zhan}}, \bibinfo {author} {\bibfnamefont {Z.}\ \bibnamefont {Bian}}, \bibinfo {author} {\bibfnamefont {J.}\ \bibnamefont {Li}}, \bibinfo {author} {\bibfnamefont {Y.}\ \bibnamefont {Zhang}}, \ and\ \bibinfo {author} {\bibfnamefont {P.}\ \bibnamefont {Xue}},\ }\href {\doibase 10.1103/PhysRevA.93.052108} {\bibfield {journal} {\bibinfo {journal} {Phys. Rev. A}\ }\textbf {\bibinfo {volume} {93}},\ \bibinfo {pages} {052108} (\bibinfo {year} {2016})}\BibitemShut {NoStop}%
  \bibitem [{\citenamefont {Maziero}\ \emph {et~al.}(2017)\citenamefont {Maziero}, \citenamefont {Maziero}}]{Maziero} \BibitemOpen \bibfield {author} {\bibinfo {author} {\bibfnamefont {J.}~\bibnamefont {Maziero}}, \bibinfo {author} \ }\href {\doibase 10.1590/1806-9126-rbef-2017-0014} {\bibfield {journal} {\bibinfo {journal} {Rev. Bras. Ens. Fis.}\ }\textbf {\bibinfo {volume} {39}},\ \bibinfo {pages} {e4306} (\bibinfo {year} {2017})}\BibitemShut {NoStop}%
  \bibitem [{\citenamefont {Verstraete}\ \emph {et~al.}(2001)\citenamefont
  {Verstraete}, \citenamefont {Audenaert},\ and\ \citenamefont
  {De~Moor}}]{mems}%
  \BibitemOpen
  \bibfield  {author} {\bibinfo {author} {\bibfnamefont {F.}~\bibnamefont
  {Verstraete}}, \bibinfo {author} {\bibfnamefont {K.}~\bibnamefont
  {Audenaert}}, \ and\ \bibinfo {author} {\bibfnamefont {B.}~\bibnamefont
  {De~Moor}},\ }\href {\doibase 10.1103/PhysRevA.64.012316} {\bibfield
  {journal} {\bibinfo  {journal} {Phys. Rev. A}\ }\textbf {\bibinfo {volume}
  {64}},\ \bibinfo {pages} {012316} (\bibinfo {year} {2001})}\BibitemShut
  {NoStop}%
\bibitem [{\citenamefont {Cheng}\ and\ \citenamefont {Hall}(2015)}]{Hall}%
  \BibitemOpen
  \bibfield  {author} {\bibinfo {author} {\bibfnamefont {S.}~\bibnamefont
  {Cheng}}\ and\ \bibinfo {author} {\bibfnamefont {M.~J.~W.}\ \bibnamefont
  {Hall}},\ }\href {\doibase 10.1103/PhysRevA.92.042101} {\bibfield  {journal}
  {\bibinfo  {journal} {Phys. Rev. A}\ }\textbf {\bibinfo {volume} {92}},\
  \bibinfo {pages} {042101} (\bibinfo {year} {2015})}\BibitemShut {NoStop}%
\bibitem [{\citenamefont {Yao}\ \emph {et~al.}(2016)\citenamefont {Yao},
  \citenamefont {Dong}, \citenamefont {Ge}, \citenamefont {Li},\ and\
  \citenamefont {Sun}}]{Yao}%
  \BibitemOpen
  \bibfield  {author} {\bibinfo {author} {\bibfnamefont {Y.}~\bibnamefont
  {Yao}}, \bibinfo {author} {\bibfnamefont {G.~H.}\ \bibnamefont {Dong}},
  \bibinfo {author} {\bibfnamefont {L.}~\bibnamefont {Ge}}, \bibinfo {author}
  {\bibfnamefont {M.}~\bibnamefont {Li}}, \ and\ \bibinfo {author}
  {\bibfnamefont {C.~P.}\ \bibnamefont {Sun}},\ }\href {\doibase
  10.1103/PhysRevA.94.062339} {\bibfield  {journal} {\bibinfo  {journal} {Phys.
  Rev. A}\ }\textbf {\bibinfo {volume} {94}},\ \bibinfo {pages} {062339}
  (\bibinfo {year} {2016})}\BibitemShut {NoStop}%
\end{thebibliography}%

\end{document}